\documentclass[aps,prb,reprint]{revtex4-1}
\usepackage{amsfonts,amssymb,amsmath,graphicx,color}
\usepackage[colorlinks=true,citecolor=blue,linkcolor=blue,urlcolor=blue]{hyperref}
\usepackage{ulem}
\allowdisplaybreaks[4]
\newcommand\dv[1]{\dot {\vec #1}}
\newcommand\tc[1]{\tilde {\cal #1}}
\newcommand\cc{{\rm c.c.}}
\DeclareMathOperator\tr{tr}
\begin{document}
\title{Gradient expansion formalism for generic spin torques}
\author{Atsuo Shitade}
\affiliation{RIKEN Center for Emergent Matter Science, 2-1 Hirosawa, Wako, Saitama 351-0198, Japan}
\date{\today}
\begin{abstract}
  We propose a new quantum-mechanical formalism to calculate spin torques based on the gradient expansion,
  which naturally involves spacetime gradients of the magnetization and electromagnetic fields.
  We have no assumption in the small-amplitude formalism or no difficulty in the SU($2$) gauge transformation formalism.
  As a representative, we calculate the spin renormalization, Gilbert damping, spin-transfer torque, and $\beta$-term
  in a three-dimensional ferromagnetic metal with nonmagnetic and magnetic impurities being taken into account within the self-consistent Born approximation.
  Our results serve as a first-principles formalism for spin torques.
\end{abstract}
\maketitle
\section{Introduction} \label{sec:introduction}
Spin torques have been investigated both theoretically and experimentally in the field of magnetic spintronics
since the celebrated discovery of the current-induced magnetization reversal by the spin transfer torque (STT)~\cite{10.1063/1.333530,10.1063/1.351045,Slonczewski1996L1,PhysRevB.54.9353,PhysRevLett.92.086601}.
When an electric field is applied to a ferromagnetic metal with magnetic structures such as domain walls and skyrmions,
the spin-polarized current flows, and electron spin is transferred to the magnetization via the exchange interaction.
Furthermore, the so-called $\beta$-term arises from spin relaxation~\cite{PhysRevLett.93.127204,0295-5075-69-6-990,PhysRevB.74.144405,JPSJ.75.113706,PhysRevB.75.174414,JPSJ.76.063710}.
Electronic contributions to spin torques in a ferromagnetic metal without spin-orbit interactions (SOIs) are expressed by
\begin{equation}
  {\vec \tau}
  = -\hbar s {\dv n} - \hbar \alpha {\vec n} \times {\dv n} - ({\vec j}_{\rm s} \cdot {\vec \partial}) {\vec n} - \beta {\vec n} \times ({\vec j}_{\rm s} \cdot {\vec \partial}) {\vec n}, \label{eq:torque}
\end{equation}
in which ${\vec n}$ is the magnetization which is dynamical and nonuniform.
$s$ and $\alpha$ are the spin renormalization and electronic contribution to the Gilbert damping, respectively.
The third and fourth terms are the STT and $\beta$-term driven by the spin-polarized current ${\vec j}_{\rm s}$.
In the presence of SOIs, another spin torque called the spin-orbit torque is allowed even without magnetic structures~\cite{PhysRevB.77.214429,PhysRevB.78.212405,PhysRevB.79.094422}.
In real materials, both magnetic structures and SOIs do exist,
and hence a systematic formalism to calculate these spin torques is desired~\cite{PhysRevB.86.094406}.

To calculate spin torques quantitatively, a quantum-mechanical formalism is desirable.
It is difficult to take into account spin relaxation systematically in the semiclassical Boltzmann theory~\cite{PhysRevB.74.144405,PhysRevB.75.174414,PhysRevB.78.212405,PhysRevB.79.094422,PhysRevB.86.094406},
and phenomenological treatment may even lead to incorrect results on the $\beta$-term~\cite{PhysRevB.74.144405}.
The small-amplitude formalism, in which small transverse fluctuations around a uniform state are assumed, is quantum-mechanical
but cannot be applied to the finite-amplitude dynamics except for simple cases without SOIs~\cite{JPSJ.75.113706}.
The SU($2$) gauge transformation formalism, where a magnetic structure is transformed to a uniform state, is also quantum-mechanical and correct~\cite{PhysRevLett.92.086601,JPSJ.76.063710,PhysRevB.77.214429}.
However, we should be careful when we deal with magnetic impurities~\cite{JPSJ.76.063710}.
Magnetic impurities become dynamical and nonuniform by the SU($2$) gauge transformation, which yields the additional SU($2$) gauge field.
If this contribution is not taken into account, the Gilbert damping vanishes.

Here we propose a new quantum-mechanical formalism to calculate generic spin torques based on the gradient expansion.
As a representative, we calculate four terms in Eq.~\eqref{eq:torque} in a three-dimensional ($3$d) ferromagnetic metal with nonmagnetic and magnetic impurities.
The gradient expansion is a perturbation theory with respect to spacetime gradients~\cite{9780521874991,9780521760829}
as well as electromagnetic fields~\cite{0953-8984-6-39-010,LEVANDA2001199,PhysRevB.82.195316}
in terms of the Wigner representations of the Keldysh Green's functions.
The former two terms in Eq.~\eqref{eq:torque} are linear responses of electron spin to a temporal gradient of the magnetization,
and the latter two are the second-order responses to a spatial gradient and an electric field.
As mentioned in Ref.~\onlinecite{PhysRevB.75.174414},
it is a natural extension of the semiclassical Boltzmann theory~\cite{PhysRevB.74.144405,PhysRevB.75.174414,PhysRevB.78.212405,PhysRevB.79.094422,PhysRevB.86.094406}.
We do not have to pay any attention to the SU($2$) gauge field even in the presence of magnetic impurities, SOIs, and sublattice degrees of freedom as in antiferromagnets.

\section{Gradient expansion} \label{sec:gradient}
In this Section, we review the gradient expansion of the Keldysh Green's function with external gauge fields being taken into account.
We do not rely on any specific form of the Hamiltonian, which may be disordered or interacting.
Furthermore, gauge fields may be abelian or nonabelian.
The gradient expansion was already carried out up to the infinite order in the absence of gauge fields~\cite{9780521874991,9780521760829} and in the abelian case~\cite{0953-8984-6-39-010,LEVANDA2001199}
and up to the first order in the nonabelian case~\cite{PhysRevB.82.195316}.
Although we are interested in the abelian case,
we give rigorous derivation up to the fourth order in the nonabelian case with the help of the nonabelian Stokes theorem~\cite{Arefeva1980,PhysRevD.22.3090}.

\subsection{Locally covariant Keldysh Green's function} \label{sub:covariant}
When we carry out the gradient expansion, it is essential to keep the local gauge covariance.
First, let us explain its meaning here.
Under a gauge transformation $\psi^{\prime}(x) = V(x) \psi(x)$ for a field $\psi(x)$,
gauge fields ${\cal A}_{\mu}(x)$, a locally gauge-covariant quantity ${\tilde A}(x)$, and the Keldysh Green's function ${\hat G}(x_1, x_2)$ transform as
\begin{subequations} \begin{align}
  {\cal A}^{\prime}_{\mu}(x)
  = & V(x) {\cal A}_{\mu}(x) V^{\dag}(x) - i \hbar [\partial_{\mu} V(x)] V^{\dag}(x), \label{eq:gauge1a} \\
  {\tilde A}^{\prime}(x)
  = & V(x) {\tilde A}(x) V^{\dag}(x), \label{eq:gauge1b} \\
  {\hat G}^{\prime}(x_1, x_2)
  = & V(x_1) {\hat G}(x_1, x_2) V^{\dag}(x_2). \label{eq:gauge1c}
\end{align} \label{eq:gauge1}\end{subequations}
The Green's function ${\hat G}(x_1, x_2)$ with the hat symbol is gauge-covariant in the sense of Eq.~\eqref{eq:gauge1c}.
However, in the Wigner representation defined later in Eq.~\eqref{eq:wigner1},
the center-of-mass coordinate $X_{12} \equiv (x_1 + x_2)/2$ is the only coordinate,
and hence the Green's function should be defined as locally gauge-covariant with respect to $X_{12}$.
It can be achieved by introducing the Wilson line,
\begin{equation}
  W(x_1, x_2)
  \equiv P \exp \left[-\frac{1}{i \hbar} \int_{x_2}^{x_1} {\rm d} y^{\mu} {\cal A}_{\mu}(y)\right], \label{eq:wilson1}
\end{equation}
which transforms in the same way as the Green's function, i.e., $W^{\prime}(x_1, x_2) = V(x_1) W(x_1, x_2) V^{\dag}(x_2)$.
$P$ is the path-ordered product.
The locally gauge-covariant Green's function ${\tilde G}(x_1, x_2)$ with the tilde symbol is then defined by~\cite{0953-8984-6-39-010,LEVANDA2001199,PhysRevB.82.195316}
\begin{equation}
  {\tilde G}(x_1, x_2)
  \equiv W(X_{12}, x_1) {\hat G}(x_1, x_2) W(x_2, X_{12}), \label{eq:wilson2}
\end{equation}
which transforms as
\begin{equation}
  {\tilde G}^{\prime}(x_1, x_2)
  = V(X_{12}) {\tilde G}(x_1, x_2) V^{\dag}(X_{12}), \label{eq:wilson3}
\end{equation}
instead of Eq.~\eqref{eq:gauge1c}.
Similarly to Eq.~\eqref{eq:wilson2}, all the two-point quantities with the hat symbol should be replaced by those with the tilde symbol.

\subsection{Gauge-covariant Wigner representation} \label{sub:wigner}
Next, we define the Wigner representation of the locally gauge-covariant Green's function~\cite{0953-8984-6-39-010,LEVANDA2001199,PhysRevB.82.195316},
\begin{equation}
  {\tilde G}(X_{12}, p_{12})
  \equiv \int {\rm d}^D x_{12} e^{p_{12 \mu} x_{12}^{\mu}/i \hbar} {\tilde G}(x_1, x_2), \label{eq:wigner1}
\end{equation}
where $X_{12} \equiv (x_1 + x_2)/2$ and $x_{12} \equiv x_1 - x_2$ are the center-of-mass and relative coordinates, respectively, and $p_{12}$ is the relative momentum.
$D$ is the spacetime dimension.
Dynamics of the Green's function is determined by the Dyson equation involving convolution,
which is a two-point quantity defined by
\begin{equation}
  \widehat{A \ast B}(x_1, x_2)
  \equiv \int {\rm d}^D x_3 {\hat A}(x_1, x_3) {\hat B}(x_3, x_2), \label{eq:conv}
\end{equation}
for any two-point quantities ${\hat A}$ and ${\hat B}$.
Therefore, we have to find the Wigner representation of the locally gauge-covariant convolution,
\begin{equation}
  {\tilde A}(X_{12}, p_{12}) \star {\tilde B}(X_{12}, p_{12})
  \equiv \widetilde{A \ast B}(X_{12}, p_{12}). \label{eq:moyal1}
\end{equation}
Since the Wigner representation is just the Fourier transformation with respect to $x_{12}$,
convolution turns into the simple product ${\tilde A}(p_{12}) {\tilde B}(p_{12})$ for a translationally invariant system in the absence of gauge fields;
otherwise, it becomes noncommutative and is called the Moyal product.
It is evaluated by expanding Eq.~\eqref{eq:moyal1} with respect to the relative coordinates $x_{13}$ and $x_{32}$ as in Appendix~\ref{app:derivation} and is expressed by
\begin{widetext}
\begin{subequations} \begin{align}
  {\tilde A} \star {\tilde B}
  = & {\tilde A} {\tilde B} + (i \hbar/2) {\cal P}_D({\tilde A}, {\tilde B}) + (i \hbar/2) {\cal P}_{\cal F}({\tilde A}, {\tilde B}) \notag \\
  & + (1/2!) (i \hbar/2)^2 {\cal P}_{D^2}({\tilde A}, {\tilde B}) + (i \hbar/2)^2 {\cal P}_{D \ast {\cal F}}({\tilde A}, {\tilde B})
  + (1/2!) (i \hbar/2)^2 {\cal P}_{{\cal F}^2}({\tilde A}, {\tilde B}), \label{eq:moyal2a} \\
  {\cal P}_D({\tilde A}, {\tilde B})
  \equiv & D_{X^{\lambda}} {\tilde A} \partial_{p_{\lambda}} {\tilde B} - \partial_{p_{\lambda}} {\tilde A} D_{X^{\lambda}} {\tilde B}, \label{eq:moyal2b} \\
  {\cal P}_{\cal F}({\tilde A}, {\tilde B})
  \equiv & ({\cal F}_{\mu \nu} \partial_{p_{\mu}} {\tilde A} \partial_{p_{\nu}} {\tilde B}
  + 2 \partial_{p_{\mu}} {\tilde A} {\cal F}_{\mu \nu} \partial_{p_{\nu}} {\tilde B}
  + \partial_{p_{\mu}} {\tilde A} \partial_{p_{\nu}} {\tilde B} {\cal F}_{\mu \nu})/4, \label{eq:moyal2c} \\
  {\cal P}_{D^2}({\tilde A}, {\tilde B})
  \equiv & D_{X^{\lambda_1}} D_{X^{\lambda_2}} {\tilde A} \partial_{p_{\lambda_1}} \partial_{p_{\lambda_2}} {\tilde B}
  - 2 D_{X^{\lambda_1}} \partial_{p_{\lambda_2}} {\tilde A} \partial_{p_{\lambda_1}} D_{X^{\lambda_2}} {\tilde B}
  + \partial_{p_{\lambda_1}} \partial_{p_{\lambda_2}} {\tilde A} D_{X^{\lambda_1}} D_{X^{\lambda_2}} {\tilde B}, \label{eq:moyal2d} \\
  {\cal P}_{D \ast {\cal F}}({\tilde A}, {\tilde B})
  \equiv & [{\cal F}_{\mu \nu} (D_{X^{\lambda}} \partial_{p_{\mu}} {\tilde A} \partial_{p_{\lambda}} \partial_{p_{\nu}} {\tilde B}
  - \partial_{p_{\lambda}} \partial_{p_{\mu}} {\tilde A} D_{X^{\lambda}} \partial_{p_{\nu}} {\tilde B}) \notag \\
  & + 2 (D_{X^{\lambda}} \partial_{p_{\mu}} {\tilde A} {\cal F}_{\mu \nu} \partial_{p_{\lambda}} \partial_{p_{\nu}} {\tilde B}
  - \partial_{p_{\lambda}} \partial_{p_{\mu}} {\tilde A} {\cal F}_{\mu \nu} D_{X^{\lambda}} \partial_{p_{\nu}} {\tilde B}) \notag \\
  & + (D_{X^{\lambda}} \partial_{p_{\mu}} {\tilde A} \partial_{p_{\lambda}} \partial_{p_{\nu}} {\tilde B}
  - \partial_{p_{\lambda}} \partial_{p_{\mu}} {\tilde A} D_{X^{\lambda}} \partial_{p_{\nu}} {\tilde B}) {\cal F}_{\mu \nu}]/4, \label{eq:moyal2e} \\
  {\cal P}_{{\cal F}^2}({\tilde A}, {\tilde B})
  \equiv & ({\cal F}_{\mu_1 \nu_1} {\cal F}_{\mu_2 \nu_2} \partial_{p_{\mu_1}} \partial_{p_{\mu_2}} {\tilde A} \partial_{p_{\nu_1}} \partial_{p_{\nu_2}} {\tilde B}
  + 4 \partial_{p_{\mu_1}} \partial_{p_{\mu_2}} {\tilde A} {\cal F}_{\mu_1 \nu_1} {\cal F}_{\mu_2 \nu_2} \partial_{p_{\nu_1}} \partial_{p_{\nu_2}} {\tilde B} \notag \\
  & + \partial_{p_{\mu_1}} \partial_{p_{\mu_2}} {\tilde A} \partial_{p_{\nu_1}} \partial_{p_{\nu_2}} {\tilde B} {\cal F}_{\mu_1 \nu_1} {\cal F}_{\mu_2 \nu_2}
  + 4 {\cal F}_{\mu_1 \nu_1} \partial_{p_{\mu_1}} \partial_{p_{\mu_2}} {\tilde A} {\cal F}_{\mu_2 \nu_2} \partial_{p_{\nu_1}} \partial_{p_{\nu_2}} {\tilde B} \notag \\
  & + 4 \partial_{p_{\mu_1}} \partial_{p_{\mu_2}} {\tilde A} {\cal F}_{\mu_1 \nu_1} \partial_{p_{\nu_1}} \partial_{p_{\nu_2}} {\tilde B} {\cal F}_{\mu_2 \nu_2}
  + 2 {\cal F}_{\mu_1 \nu_1} \partial_{p_{\mu_1}} \partial_{p_{\mu_2}} {\tilde A} \partial_{p_{\nu_1}} \partial_{p_{\nu_2}} {\tilde B} {\cal F}_{\mu_2 \nu_2})/4^2. \label{eq:moyal2f}
\end{align} \label{eq:moyal2}\end{subequations}
\end{widetext}
Here a covariant derivative and a field strength are defined by
\begin{subequations} \begin{align}
  D_{X^{\mu}} {\tilde A}(X, p)
  \equiv & \partial_{X^{\mu}} {\tilde A}(X, p) \notag \\
  & + [{\cal A}_{\mu}(X), {\tilde A}(X, p)]/i \hbar, \label{eq:gauge2a} \\
  {\cal F}_{\mu \nu}(X)
  \equiv & \partial_{X^{\mu}} {\cal A}_{\nu}(X) - \partial_{X^{\nu}} {\cal A}_{\mu}(X) \notag \\
  & + [{\cal A}_{\mu}(X), {\cal A}_{\nu}(X)]/i \hbar. \label{eq:gauge2b}
\end{align} \label{eq:gauge2}\end{subequations}
For simplicity, the arguments $X, p$ are omitted in Eq.~\eqref{eq:moyal2} and below.

After all, the Moyal product is regarded as a perturbation theory with respect to spacetime gradients as well as field strengths, but not to gauge fields.
Thus, the gauge covariance of the results is guaranteed.
${\cal P}_D$ and ${\cal P}_{\cal F}$ denote the first-order contributions with respect to spacetime gradients $D$ and field strengths ${\cal F}$, respectively.
${\cal P}_{D \ast {\cal F}}$ is the mixed second-order contribution involving $D$ and ${\cal F}$.
We also write down the second order with respect to ${\cal F}$ in Eq.~\eqref{eq:moyal2f}, which may be useful for studying other nonlinear responses in the future.
In order to derive ${\cal P}_{{\cal F}^2}$, we need the fourth order with respect to $x_{13}$ and $x_{32}$ and obtain many other terms.
All the terms up to the fourth order are written in Eq.~\eqref{eq:wigner8}.

\subsection{Gradient expansion up to the second order} \label{sub:gradient}
Here we derive the gradient expansion of the Keldysh Green's function.
We focus on the abelian case and assume a static and uniform field strength.
Similarly to Eq.~\eqref{eq:moyal2a}, we expand the Green's function and self-energy as~\cite{PTP.116.61,Shitade01122014}
\begin{subequations} \begin{align}
  {\tilde G}
  = & {\tilde G}_0 + (\hbar/2) {\tilde G}_D + (\hbar/2) {\tilde G}_{\cal F} + (1/2!) (\hbar/2)^2 {\tilde G}_{D^2} \notag \\
  & + (\hbar/2)^2 {\tilde G}_{D \ast {\cal F}} + (1/2!) (\hbar/2)^2 {\tilde G}_{{\cal F}^2}, \label{eq:grad1a} \\
  {\tilde \Sigma}
  = & {\tilde \Sigma}_0 + (\hbar/2) {\tilde \Sigma}_D + (\hbar/2) {\tilde \Sigma}_{\cal F} + (1/2!) (\hbar/2)^2 {\tilde \Sigma}_{D^2} \notag \\
  & + (\hbar/2)^2 {\tilde \Sigma}_{D \ast {\cal F}} + (1/2!) (\hbar/2)^2 {\tilde \Sigma}_{{\cal F}^2}. \label{eq:grad1b}
\end{align} \label{eq:grad1}\end{subequations}
Note that ${\tilde G}_0$ is the unperturbed Green's function with disorder or interactions being taken into account.
${\tilde G}_P$ and ${\tilde G}_{P \ast Q}$ ($P, Q = D, {\cal F}$) are the first and second orders with respect to spacetime gradients or field strengths, respectively.
By substituting these into the left Dyson equation,
\begin{equation}
  ({\tc L} - {\tilde \Sigma}) \star {\tilde G}
  = 1, \label{eq:dyson1}
\end{equation}
in which ${\tc L}$ is the Lagrangian,
we get ${\tilde G}_0 = ({\tc L} - {\tilde \Sigma}_0)^{-1}$ and
\begin{subequations} \begin{align}
  {\tilde G}_0^{-1} {\tilde G}_P
  = & {\tilde \Sigma}_P {\tilde G}_0 - i {\cal P}_P({\tilde G}_0^{-1}, {\tilde G}_0), \label{eq:grad2a} \\
  {\tilde G}_0^{-1} {\tilde G}_{P \ast Q}
  = & {\tilde \Sigma}_{P \ast Q} {\tilde G}_0 - i^2 {\cal P}_{P \ast Q}({\tilde G}_0^{-1}, {\tilde G}_0) \notag \\
  & + [{\tilde \Sigma}_Q {\tilde G}_P + i {\cal P}_P({\tilde \Sigma}_Q, {\tilde G}_0) - i {\cal P}_P({\tilde G}_0^{-1}, {\tilde G}_Q) \notag \\
  & + (P \leftrightarrow Q)]. \label{eq:grad2b}
\end{align} \label{eq:grad2}\end{subequations}
The self-energies are determined self-consistently.

To calculate the expectation values, the lesser Green's function is necessary.
In the real-time representation, the Green's function and self-energy are of matrix forms~\cite{PTP.116.61,Shitade01122014},
\begin{subequations} \begin{align}
  {\tilde G}
  = & \begin{bmatrix}
    G^{\rm R} & 2 G^< \\
    0 & G^{\rm A}
  \end{bmatrix}, \label{eq:keldysh1a} \\
  {\tilde \Sigma}
  = & \begin{bmatrix}
    \Sigma^{\rm R} & 2 \Sigma^< \\
    0 & \Sigma^{\rm A}
  \end{bmatrix}, \label{eq:keldysh1b}
\end{align} \label{eq:keldysh1}\end{subequations}
in which ${\rm R}$, ${\rm A}$, and $<$ indicate the retarded, advanced, and lesser components, respectively.
For the first order, the lesser one can be written as
\begin{subequations} \begin{align}
  G_P^<
  = & \pm [(G_P^{\rm R} - G_P^{\rm A}) f(-p_0) + G_P^{< (1)} f^{\prime}(-p_0)], \label{eq:first1a} \\
  \Sigma_P^<
  = & \pm [(\Sigma_P^{\rm R} - \Sigma_P^{\rm A}) f(-p_0) + \Sigma_P^{< (1)} f^{\prime}(-p_0)], \label{eq:first1b}
\end{align} \label{eq:first1}\end{subequations}
in which the upper and lower signs indicate boson and fermion, respectively, and $f(\xi) = (e^{\xi/T} \mp 1)^{-1}$ is the distribution function at temperature $T$.
By introducing ${\cal P}_P({\tilde A}, {\tilde B}) \equiv \eta_P^{IJ} \partial_I {\tilde A} \partial_J {\tilde B}$ ($I, J = X^{\mu}, p_{\nu}$) with
$\eta_D^{X^{\mu} p_{\nu}} = -\eta_D^{p_{\nu} X^{\mu}} = \delta^{\mu}_{\nu}$ and $\eta_{\cal F}^{p_{\mu} p_{\nu}} = {\cal F}_{\mu \nu}$,
we obtain an equivalent form of Eq.~\eqref{eq:grad2a}~\cite{PTP.116.61,Shitade01122014},
\begin{subequations} \begin{align}
  (G_0^{\rm R})^{-1} G_P^{\rm R}
  = & \Sigma_P^{\rm R} G_0^{\rm R} - i \eta_P^{IJ} \partial_I (G_0^{\rm R})^{-1} \partial_J G_0^{\rm R}, \label{eq:first2a} \\
  (G_0^{\rm A})^{-1} G_P^{\rm A}
  = & \Sigma_P^{\rm A} G_0^{\rm A} - i \eta_P^{IJ} \partial_I (G_0^{\rm A})^{-1} \partial_J G_0^{\rm A}, \label{eq:first2b} \\
  (G_0^{\rm R})^{-1} G_P^{< (1)}
  = & \Sigma_P^{< (1)} G_0^{\rm A}
  + i \eta_P^{I p_0} \{\partial_I (G_0^{\rm R})^{-1} (G_0^{\rm R} - G_0^{\rm A}) \notag \\
  & - [(G_0^{\rm R})^{-1} - (G_0^{\rm A})^{-1}] \partial_I G_0^{\rm A}\}. \label{eq:first2c}
\end{align} \label{eq:first2}\end{subequations}
Similarly, for the second order, we obtain the lesser Green's function,
\begin{subequations} \begin{align}
  G_{P \ast Q}^<
  = & \pm [(G_{P \ast Q}^{\rm R} - G_{P \ast Q}^{\rm A}) f(-p_0) + G_{P \ast Q}^{< (1)} f^{\prime}(-p_0) \notag \\
  & + G_{P \ast Q}^{< (2)} f^{\prime \prime}(-p_0)], \label{eq:second1a} \\
  \Sigma_{P \ast Q}^<
  = & \pm [(\Sigma_{P \ast Q}^{\rm R} - \Sigma_{P \ast Q}^{\rm A}) f(-p_0) + \Sigma_{P \ast Q}^{< (1)} f^{\prime}(-p_0) \notag \\
  & + \Sigma_{P \ast Q}^{< (2)} f^{\prime \prime}(-p_0)], \label{eq:second1b}
\end{align} \label{eq:second1}\end{subequations}
and an equivalent form of Eq.~\eqref{eq:grad2b},
\begin{widetext}
\begin{subequations} \begin{align}
  (G_0^{\rm R})^{-1} G_{P \ast Q}^{\rm R}
  = & \Sigma_{P \ast Q}^{\rm R} G_0^{\rm R}
  + [\Sigma_Q^{\rm R} G_P^{\rm R} + i \eta_P^{IJ} \partial_I \Sigma_Q^{\rm R} \partial_J G_0^{\rm R} - i \eta_P^{IJ} \partial_I (G_0^{\rm R})^{-1} \partial_J G_Q^{\rm R} + (P \leftrightarrow Q)] \notag \\
  & + \eta_P^{IJ} \eta_Q^{KL} \partial_I \partial_K (G_0^{\rm R})^{-1} \partial_J \partial_L G_0^{\rm R}, \label{eq:second2a} \\
  (G_0^{\rm A})^{-1} G_{P \ast Q}^{\rm A}
  = & \Sigma_{P \ast Q}^{\rm A} G_0^{\rm A}
  + [\Sigma_Q^{\rm A} G_P^{\rm A} + i \eta_P^{IJ} \partial_I \Sigma_Q^{\rm A} \partial_J G_0^{\rm A} - i \eta_P^{IJ} \partial_I (G_0^{\rm A})^{-1} \partial_J G_Q^{\rm A} + (P \leftrightarrow Q)] \notag \\
  & + \eta_P^{IJ} \eta_Q^{KL} \partial_I \partial_K (G_0^{\rm A})^{-1} \partial_J \partial_L G_0^{\rm A}, \label{eq:second2b} \\
  (G_0^{\rm R})^{-1} G_{P \ast Q}^{< (1)}
  = & \Sigma_{P \ast Q}^{< (1)} G_0^{\rm A} + \left(\Sigma_Q^{\rm R} G_P^{< (1)} + \Sigma_Q^{< (1)} G_P^{\rm A}
  + i \eta_P^{IJ} \partial_I \Sigma_Q^{< (1)} \partial_J G_0^{\rm A}
  - i \eta_P^{I p_0} [\partial_I \Sigma_Q^{\rm R} (G_0^{\rm R} - G_0^{\rm A}) - (\Sigma_Q^{\rm R} - \Sigma_Q^{\rm A}) \partial_I G_0^{\rm A}]\right. \notag \\
  & - i \eta_P^{IJ} \partial_I (G_0^{\rm R})^{-1} \partial_J G_Q^{< (1)}
  + i \eta_P^{I p_0} \{\partial_I (G_0^{\rm R})^{-1} (G_Q^{\rm R} - G_Q^{\rm A}) - [(G_0^{\rm R})^{-1} - (G_0^{\rm A})^{-1}] \partial_I G_Q^{\rm A}\} \notag \\
  & \left.- \eta_P^{I p_0} \eta_Q^{KL} \{\partial_I \partial_K (G_0^{\rm R})^{-1} \partial_L (G_0^{\rm R} - G_0^{\rm A}) + \partial_L [(G_0^{\rm R})^{-1} - (G_0^{\rm A})^{-1}] \partial_I \partial_K G_0^{\rm A}\}
  + (P \leftrightarrow Q)\right), \label{eq:second2c} \\
  (G_0^{\rm R})^{-1} G_{P \ast Q}^{< (2)}
  = & \Sigma_{P \ast Q}^{< (2)} G_0^{\rm A}
  + [i \eta_P^{I p_0} \Sigma_Q^{< (1)} \partial_I G_0^{\rm A} + i \eta_P^{I p_0} \partial_I (G_0^{\rm R})^{-1} G_Q^{< (1)} + (P \leftrightarrow Q)] \notag \\
  & + \eta_P^{I p_0} \eta_Q^{K p_0} \{\partial_I \partial_K (G_0^{\rm R})^{-1} (G_0^{\rm R} - G_0^{\rm A}) + [(G_0^{\rm R})^{-1} - (G_0^{\rm A})^{-1}] \partial_I \partial_K G_0^{\rm A}\}. \label{eq:second2d}
\end{align} \label{eq:second2}\end{subequations}
\end{widetext}
Note that the left and right Dyson equations are equivalent as explicitly proved in Appendix~\ref{app:equivalence}.
Generally, the $n$th-order lesser Green's function with respect to a temporal gradient or an electric field involves the $n$th derivative of the distribution function.

\subsection{Spin torques}
Spin torques are proportional to the spin expectation value.
The spin expectation value is given by
\begin{align}
  \langle {\vec \sigma} \rangle
  = & \pm i \hbar \int \frac{{\rm d}^D p}{(2 \pi \hbar)^D} \tr {\vec \sigma} G^< \notag \\
  = & \pm i \hbar \int \frac{{\rm d}^D p}{(2 \pi \hbar)^D} \tr {\vec \sigma} [G_0^< + (\hbar/2) G_D^< + (\hbar/2) G_{\cal F}^< \notag \\
  & + (1/2!) (\hbar/2)^2 G_{D^2}^< + (\hbar/2)^2 G_{D \ast {\cal F}}^< \notag \\
  & + (1/2!) (\hbar/2)^2 G_{{\cal F}^2}^<]. \label{eq:spin}
\end{align}
Among many terms in Eq.~\eqref{eq:spin}, $G_D^<$ yields the spin renormalization and Gilbert damping.
$G_{\cal F}^<$ yields the spin-orbit torque in the presence of SOIs.
In order to calculate the STT and $\beta$-term driven by an electric field, $G_{D \ast {\cal F}}^<$ is necessary.
Equations~\eqref{eq:first1}-\eqref{eq:spin} are our central results for calculating spin torques in generic systems.

\section{Application to a $3$d ferromagnetic metal} \label{sec:torque}
As an example, we explicitly calculate spin torques in a $3$d ferromagnetic metal,
\begin{equation}
  {\cal H}(X, {\vec p})
  = {\vec p}^2/2 m - \mu - J {\vec n}(X) \cdot {\vec \sigma}, \label{eq:ferro}
\end{equation}
with the chemical potential $\mu$ and $[{\vec n}(X)]^2 = 1$.
We take into account nonmagnetic and magnetic impurities,
\begin{align}
  V_{\rm imp}({\vec X})
  = & \sum_j^{N_{\rm i}} v_{\rm i} \delta({\vec X} - {\vec X}_{{\rm i} j}) \notag \\
  & + \sum_j^{N_{\rm s}} v_{\rm s} {\vec m}_j \cdot {\vec \sigma} \delta({\vec X} - {\vec X}_{{\rm s} j}). \label{eq:imp}
\end{align}
Magnetic impurities are assumed to be isotropic, namely, $\overline{m_i^a} = 0, \overline{m_i^a m_j^b} = \delta^{ab}/3$ after average over the magnetization directions.
The same system was studied within the Born approximation in the literature~\cite{JPSJ.75.113706,JPSJ.76.063710}.
Here we employ the self-consistent Born approximation, but it does not change any results quantitatively.

We only have to calculate the momentum integrals of $G_0^<$, $G_D^<$, $G_{\cal F}^<$, and $G_{D \ast {\cal F}}^<$ with Eqs.~\eqref{eq:first1}-\eqref{eq:second2} in order.
First, the unperturbed Green's function is given by
\begin{subequations} \begin{align}
  (G_0^{\rm R})^{-1}(X, \xi, {\vec p})
  = & \xi - {\cal H}(X, {\vec p}) - \Sigma_0^{\rm R}(X, \xi) \notag \\
  = & \xi^{\rm R}(\xi) - x + J^{\rm R}(\xi) {\vec n}(X) \cdot {\vec \sigma}, \label{eq:ferro01a} \\
  G_0^{\rm R}(X, \xi, {\vec p})
  = & \Gamma_0^{\rm R}(\xi, {\vec p}) [\xi^{\rm R}(\xi) - x \notag \\
  & - J^{\rm R}(\xi) {\vec n}(X) \cdot {\vec \sigma}], \label{eq:ferro01b} \\
  \Gamma_0^{\rm R}(\xi, {\vec p})
  = & \{[\xi^{\rm R}(\xi) - x]^2 - [J^{\rm R}(\xi)]^2\}^{-1}, \label{eq:ferro01c}
\end{align} \label{eq:ferro01}\end{subequations}
with $\xi^{\rm R}(\xi) \equiv \xi + \mu - \Sigma_{00}^{\rm R}(\xi)$, $J^{\rm R}(\xi) \equiv J - \Sigma_{03}^{\rm R}(\xi)$, and $x \equiv {\vec p}^2/2 m$.
Note that the self-energy is expressed as $\Sigma_0^{\rm R}(X, \xi) = \Sigma_{00}^{\rm R}(\xi) + \Sigma_{03}^{\rm R}(\xi) {\vec n}(X) \cdot {\vec \sigma}$ due to the spin rotation symmetry.
For convenience, we define $g$ as the momentum integral of $G$ mutiplied by $4 \pi (\hbar^2/2 m)^{3/2}$
and introduce $\gamma_{\rm i} \equiv n_{\rm i} v_{\rm i}^2 (2 m/\hbar^2)^{3/2}/4 \pi, \gamma_{\rm s} \equiv n_{\rm s} v_{\rm s}^2 (2 m/\hbar^2)^{3/2}/4 \pi$.
The momentum integral and self-energy are obtained by self-consistently solving
\begin{subequations} \begin{align}
  g_0^{\rm R}(X, \xi)
  \equiv & 4 \pi \left(\frac{\hbar^2}{2 m}\right)^{3/2} \int \frac{{\rm d}^3 p}{(2 \pi \hbar)^3} G_0^{\rm R}(X, \xi, {\vec p}) \notag \\
  = & I_{11}^{\rm R}(\xi) - I_{01}^{\rm R}(\xi) {\vec n}(X) \cdot {\vec \sigma}, \label{eq:ferro02a} \\
  \Sigma_0^{\rm R}(X, \xi)
  = & \gamma_{\rm i} g_0^{\rm R}(X, \xi) + \gamma_{\rm s} \overline{{\vec m}_i \cdot {\vec \sigma} g_0^{\rm R}(X, \xi) {\vec m}_j \cdot {\vec \sigma}} \notag \\
  = & (\gamma_{\rm i} + \gamma_{\rm s}) g_{00}^{\rm R}(\xi) \notag \\
  + & (\gamma_{\rm i} - \gamma_{\rm s}/3) g_{03}^{\rm R}(\xi) {\vec n}(X) \cdot {\vec \sigma}, \label{eq:ferro02b}
\end{align} \label{eq:ferro02}\end{subequations}
in which we define
\begin{equation}
  I_{mn}^{\rm R}(\xi)
  \equiv \int_0^{\Lambda} \frac{\sqrt{x} {\rm d} x}{\pi} \frac{[\xi^{\rm R}(\xi) - x]^m [J^{\rm R}(\xi)]^{2 n - m - 1}}{\{[\xi^{\rm R}(\xi) - x]^2 - [J^{\rm R}(\xi)]^2\}^n}. \label{eq:ferroint1}
\end{equation}

\begin{figure*}
  \centering
  \includegraphics[clip,width=\textwidth]{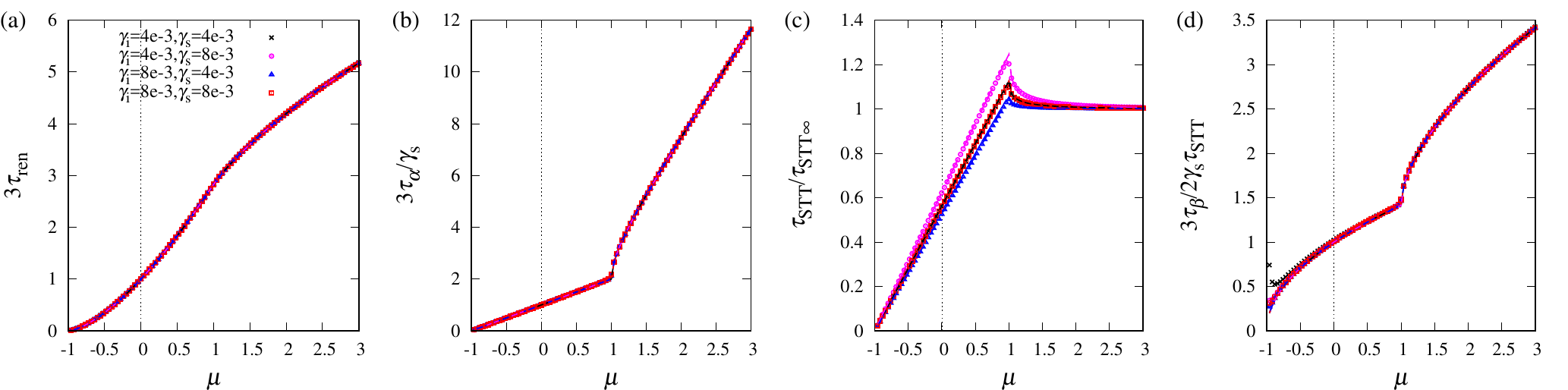}
  \caption{%
  (Color online) Chemical-potential dependences of
  (a) $3 \tau_{\rm ren}$,
  (b) $3 \tau_{\alpha}/\gamma_{\rm s}$,
  (c) $\tau_{\rm STT}/\tau_{{\rm STT} \infty}$,
  and (d) $3 \tau_{\beta}/2 \gamma_{\rm s} \tau_{\rm STT}$ for different $\gamma_{\rm i}$ and $\gamma_{\rm s}$.
  Points are obtained by the gradient expansion formalism and numerical calculation of Eqs.~\eqref{eq:torque1c}-\eqref{eq:torque1f},
  while lines are obtained by the small-amplitude~\cite{JPSJ.75.113706} and the SU($2$) gauge transformation formalisms~\cite{JPSJ.76.063710}.%
  } \label{fig:torque}
\end{figure*}

\begin{widetext}
Second, the first-order Green's functions are given by
\begin{subequations} \begin{align}
  G_D^{\rm R}(X, \xi, {\vec p})
  = & G_0^{\rm R}(X, \xi, {\vec p}) \Sigma_D^{\rm R}(X, \xi) G_0^{\rm R}(X, \xi, {\vec p}) \notag \\
  & + 2 J^{\rm R}(\xi) [\Gamma_0^{\rm R}(\xi, {\vec p})]^2
  \{J^{\rm R}(\xi) \xi^{{\rm R} \prime}(\xi) - [\xi^{\rm R}(\xi) - x] J^{{\rm R} \prime}(\xi)\} {\vec n}(X) \times {\dv n}(X) \cdot {\vec \sigma} \notag \\
  & + 2 [J^{\rm R}(\xi)]^2 [\Gamma_0^{\rm R}(\xi, {\vec p})]^2 (p^i/m) {\vec n}(X) \times \partial_{X^i} {\vec n}(X) \cdot \sigma, \label{eq:ferro11a} \\
  G_D^{< (1)}(X, \xi, {\vec p})
  = & G_0^{\rm R}(X, \xi, {\vec p}) \Sigma_D^{< (1)}(X, \xi) G_0^{\rm A}(X, \xi, {\vec p}) \notag \\
  & - i [J^{\rm R}(\xi) + J^{\rm A}(\xi)] |\Gamma_0^{\rm R}(\xi, {\vec p})|^2 [|\xi^{\rm R}(\xi) - x|^2 - |J^{\rm R}(\xi)|^2] {\dv n}(X) \cdot {\vec \sigma} \notag \\
  & + [J^{\rm R}(\xi) + J^{\rm A}(\xi)] |\Gamma_0^{\rm R}(\xi, {\vec p})|^2
  \{[\xi^{\rm R}(\xi) - x] J^{\rm A}(\xi) - [\xi^{\rm A}(\xi) - x] J^{\rm R}(\xi)\} {\vec n}(X) \times {\dv n}(X) \cdot {\vec \sigma} \notag \\
  & - i \partial_{X^0} [G_0^{\rm R}(X, \xi, {\vec p}) + G_0^{\rm A}(X, \xi, {\vec p})], \label{eq:ferro11b} \\
  G_{\cal F}^{\rm R}(X, \xi, {\vec p})
  = & G_0^{\rm R}(X, \xi, {\vec p}) \Sigma_{\cal F}^{\rm R}(X, \xi) G_0^{\rm R}(X, \xi, {\vec p}), \label{eq:ferro11c} \\
  G_{\cal F}^{< (1)}(X, \xi, {\vec p})
  = & G_0^{\rm R}(X, \xi, {\vec p}) \Sigma_{\cal F}^{< (1)}(X, \xi) G_0^{\rm A}(X, \xi, {\vec p}) \notag \\
  & + i {\cal F}_{j0} (p^j/m) \{2 G_0^{\rm R}(X, \xi, {\vec p}) G_0^{\rm A}(X, \xi, {\vec p}) - [G_0^{\rm R}(X, \xi, {\vec p})]^2 - [G_0^{\rm A}(X, \xi, {\vec p})]^2\}. \label{eq:ferro11d}
\end{align} \label{eq:ferro11}\end{subequations}
The self-energies are expressed as $\Sigma_D^{\rm R}(X, \xi) \equiv \Sigma_{D 2}^{\rm R}(\xi) {\vec n}(X) \times {\dv n}(X) \cdot {\vec \sigma},
\Sigma_D^{< (1)}(X, \xi) \equiv \Sigma_{D 1}^{< (1)}(\xi) {\dv n}(X) \cdot {\vec \sigma} + \Sigma_{D 2}^{< (1)}(\xi) {\vec n}(X) \times {\dv n}(X) \cdot {\vec \sigma}$
and obtained by solving the following sets of linear equations,
\begin{subequations} \begin{align}
  g_{D 2}^{\rm R}(\xi)
  = & I_{01}^{\rm R}(\xi) \Sigma_{D 2}^{\rm R}(\xi)/J^{\rm R}(\xi) + 2 [\xi^{{\rm R} \prime}(\xi) I_{02}^{\rm R}(\xi) - J^{{\rm R} \prime}(\xi) I_{12}^{\rm R}(\xi)]/J^{\rm R}(\xi), \label{eq:ferro12a} \\
  \Sigma_{D 2}^{\rm R}(\xi)
  = & (\gamma_{\rm i} - \gamma_{\rm s}/3) g_{D 2}^{\rm R}(\xi), \label{eq:ferro12b} \\
  \begin{bmatrix}
    g_{D 1}^{< (1)}(\xi) \\
    g_{D 2}^{< (1)}(\xi)
  \end{bmatrix}
  = & 
  \begin{bmatrix}
    J_1^{< (1)}(\xi) & -J_2^{< (1)}(\xi) \\
    J_2^{< (1)}(\xi) & J_1^{< (1)}(\xi)
  \end{bmatrix}  \begin{bmatrix}
    \Sigma_{D 1}^{< (1)}(\xi) \\
    \Sigma_{D 2}^{< (1)}(\xi)
  \end{bmatrix}
  - 2 i
  \begin{bmatrix}
    [J^{\rm R}(\xi) + J^{\rm A}(\xi)] J_1^{< (1)}(\xi)/2 - [I_{01}^{\rm R}(\xi) + \cc]/2 \\
    [J^{\rm R}(\xi) + J^{\rm A}(\xi)] J_2^{< (1)}(\xi)/2
  \end{bmatrix}, \label{eq:ferro12c} \\
  \begin{bmatrix}
    \Sigma_{D 1}^{< (1)}(\xi) \\
    \Sigma_{D 2}^{< (1)}(\xi)
  \end{bmatrix}
  = & (\gamma_{\rm i} - \gamma_{\rm s}/3)
  \begin{bmatrix}
    g_{D 1}^{< (1)}(\xi) \\
    g_{D 2}^{< (1)}(\xi)
  \end{bmatrix}. \label{eq:ferro12d}
\end{align} \label{eq:ferro12}\end{subequations}
Here we define
\begin{subequations} \begin{align}
  J_1^{< (1)}(\xi)
  \equiv & \int_0^{\Lambda} \frac{\sqrt{x} {\rm d} x}{\pi} \frac{|\xi^{\rm R}(\xi) - x|^2 - |J^{\rm R}(\xi)|^2}{|[\xi^{\rm R}(\xi) - x]^2 - [J^{\rm R}(\xi)]^2|^2}, \label{eq:ferroint2a} \\
  J_2^{< (1)}(\xi)
  \equiv & \int_0^{\Lambda} \frac{\sqrt{x} {\rm d} x}{\pi}
  \frac{i [\xi^{\rm R}(\xi) - x] J^{\rm A}(\xi) - i [\xi^{\rm A}(\xi) - x] J^{\rm R}(\xi)}{|[\xi^{\rm R}(\xi) - x]^2 - [J^{\rm R}(\xi)]^2|^2}, \label{eq:ferroint2b} \\
  J_3^{< (1)}(\xi)
  \equiv & J_2^{< (1)}(\xi) - \frac{2}{3} [i J^{\rm R}(\xi) - i J^{\rm A}(\xi)] \int_0^{\Lambda} \frac{\sqrt{x} {\rm d} x}{\pi}
  \frac{x}{|[\xi^{\rm R}(\xi) - x]^2 - [J^{\rm R}(\xi)]^2|^2}, \label{eq:ferroint2c} \\
  J_4^{< (1)}(\xi)
  \equiv & \int_0^{\Lambda} \frac{\sqrt{x} {\rm d} x}{\pi} \frac{2}{|[\xi^{\rm R}(\xi) - x]^2 - [J^{\rm R}(\xi)]^2|^2} \left(|J^{\rm R}(\xi)|^2\right. \notag \\
  & \left.+ \frac{2}{3} x \left\{\frac{J^{\rm R}(\xi)}{[\xi^{\rm R}(\xi) - x]^2 - [J^{\rm R}(\xi)]^2} + \frac{J^{\rm A}(\xi)}{[\xi^{\rm A}(\xi) - x]^2 - J^{{\rm A} 2}(\xi)}\right\}
  \{[\xi^{\rm R}(\xi) - x] J^{\rm A}(\xi) + [\xi^{\rm A}(\xi) - x] J^{\rm R}(\xi)\}\right). \label{eq:ferroint2d}
\end{align} \label{eq:ferroint2}\end{subequations}
These momentum integrals in Eqs.~\eqref{eq:ferroint1} and \eqref{eq:ferroint2} are explicitly calculated in Appendix~\ref{app:integral}.
The other components vanish, i.e., $\Sigma_{\cal F}^{\rm R}(X, \xi) = g_{\cal F}^{\rm R}(X, \xi) = 0$ and $\Sigma_{\cal F}^{< (1)}(X, \xi) = g_{\cal F}^{< (1)}(X, \xi) = 0$.

Third, the second-order Green's functions are given by
\begin{subequations} \begin{align}
  G_{D \ast {\cal F}}^{\rm R}({\vec X}, \xi, {\vec p})
  = & G_0^{\rm R}({\vec X}, \xi, {\vec p}) \Sigma_{D \ast {\cal F}}^{\rm R}({\vec X}, \xi) G_0^{\rm R}({\vec X}, \xi, {\vec p}) \notag \\
  & - 2 [\Gamma_0^{\rm R}(\xi, {\vec p})]^2
  \left(\{J^{\rm R}(\xi) \xi^{{\rm R} \prime}(\xi) - [\xi^{\rm R}(\xi) - x] J^{{\rm R} \prime}(\xi)\} \delta^{ij}\right. \notag \\
  & \left.- J^{{\rm R} \prime}(\xi) p^i p^j/m\right) {\cal F}_{j0} \partial_{X^i} {\vec n}({\vec X}) \cdot {\vec \sigma}/m, \label{eq:ferro21a} \\
  G_{D \ast {\cal F}}^{< (1)}({\vec X}, \xi, {\vec p})
  = & G_0^{\rm R}({\vec X}, \xi, {\vec p}) \Sigma_{D \ast {\cal F}}^{< (1)}({\vec X}, \xi) G_0^{\rm A}({\vec X}, \xi, {\vec p}) \notag \\
  & - 2 |\Gamma_0^{\rm R}(\xi, {\vec p})|^2 \left(\{[\xi^{\rm R}(\xi) - x] J^{\rm A}(\xi) - [\xi^{\rm A}(\xi) - x] J^{\rm R}(\xi)\}\delta^{ij}\right. \notag \\
  & \left.- [J^{\rm R}(\xi) - J^{\rm A}(\xi)] p^i p^j/m\right) {\cal F}_{j0} \partial_{X^i} {\vec n}({\vec X}) \cdot {\vec \sigma}/m \notag \\
  & + 4 i |\Gamma_0^{\rm R}(\xi, {\vec p})|^2 \left(|J^{\rm R}(\xi)|^2 \delta^{ij} + [\Gamma_0^{\rm R}(\xi, {\vec p}) J^{\rm R}(\xi) + \Gamma_0^{\rm A}(\xi, {\vec p}) J^{\rm A}(\xi)]\right. \notag \\
  & \left.\times \{[\xi^{\rm R}(\xi) - x] J^{\rm A}(\xi) + [\xi^{\rm A}(\xi) - x] J^{\rm R}(\xi)\} p^i p^j/m\right)
  {\cal F}_{j0} {\vec n}({\vec X}) \times \partial_{X^i} {\vec n}({\vec X}) \cdot {\vec \sigma}/m, \label{eq:ferro21b}
\end{align} \label{eq:ferro21}\end{subequations}
where we drop the $X^0$ dependence in ${\vec n}(X)$ because we are interested in the STT and $\beta$-term only.
The self-energies are expressed as $\Sigma_{D \ast {\cal F}}^{\rm R}({\vec X}, \xi) \equiv \Sigma_{D \ast {\cal F} 1}^{\rm R}(\xi) {\cal F}_{i0} \partial_{X^i} {\vec n}({\vec X}) \cdot {\vec \sigma}/m,
\Sigma_{D \ast {\cal F}}^{< (1)}({\vec X}, \xi) \equiv \Sigma_{D \ast {\cal F} 1}^{< (1)}(\xi) {\cal F}_{i0} \partial_{X^i} {\vec n}({\vec X}) \cdot {\vec \sigma}/m
+ \Sigma_{D \ast {\cal F} 2}^{< (1)}(\xi) {\cal F}_{i0} {\vec n}({\vec X}) \times \partial_{X^i} {\vec n}({\vec X}) \cdot {\vec \sigma}/m$
and obtained by solving the following sets of linear equations,
\begin{subequations} \begin{align}
  g_{D \ast {\cal F} 1}^{\rm R}(\xi)
  = & I_{01}^{\rm R}(\xi) \Sigma_{D \ast {\cal F} 1}^{\rm R}(\xi)/J^{\rm R}(\xi) \notag \\
  & - 2 \{[3 J^{\rm R}(\xi) \xi^{{\rm R} \prime}(\xi) - 2 \xi^{\rm R}(\xi) J^{{\rm R} \prime}(\xi)] I_{02}^{\rm R}(\xi)
  - J^{\rm R}(\xi) J^{{\rm R} \prime}(\xi) I_{12}^{\rm R}(\xi)\}/3 J^{{\rm R 3}}(\xi), \label{eq:ferro22a} \\
  \Sigma_{D \ast {\cal F} 1}^{\rm R}(\xi)
  = & (\gamma_{\rm i} - \gamma_{\rm s}/3) g_{D \ast {\cal F} 1}^{\rm R}(\xi), \label{eq:ferro22b} \\
  \begin{bmatrix}
    g_{D \ast {\cal F} 1}^{< (1)}(\xi) \\
    g_{D \ast {\cal F} 2}^{< (1)}(\xi)
  \end{bmatrix}
  = & 
  \begin{bmatrix}
    J_1^{< (1)}(\xi) & -J_2^{< (1)}(\xi) \\
    J_2^{< (1)}(\xi) & J_1^{< (1)}(\xi)
  \end{bmatrix}  \begin{bmatrix}
    \Sigma_{D \ast {\cal F} 1}^{< (1)}(\xi) \\
    \Sigma_{D \ast {\cal F} 2}^{< (1)}(\xi)
  \end{bmatrix}
  + 2 i
  \begin{bmatrix}
    J_3^{< (1)}(\xi) \\
    J_4^{< (1)}(\xi)
  \end{bmatrix}, \label{eq:ferro22c} \\
  \begin{bmatrix}
    \Sigma_{D \ast {\cal F} 1}^{< (1)}(\xi) \\
    \Sigma_{D \ast {\cal F} 2}^{< (1)}(\xi)
  \end{bmatrix}
  = & (\gamma_{\rm i} - \gamma_{\rm s}/3)
  \begin{bmatrix}
    g_{D \ast {\cal F} 1}^{< (1)}(\xi) \\
    g_{D \ast {\cal F} 2}^{< (1)}(\xi)
  \end{bmatrix}. \label{eq:ferro22d}
\end{align} \label{eq:ferro22}\end{subequations}

The spin expectation value is given by
\begin{subequations} \begin{align}
  \langle {\vec \sigma} \rangle(X)
  = & \pm i \hbar \int \frac{{\rm d} \xi}{2 \pi \hbar} \int \frac{{\rm d}^3 p}{(2 \pi \hbar)^3} \tr {\vec \sigma} G^<(X, \xi, {\vec p}) \notag \\
  = & \langle {\vec \sigma} \rangle_0(X)
  - \frac{\hbar}{4 \pi^2 J} \left(\frac{2 m J}{\hbar^2}\right)^{3/2} [\tau_{\alpha} - \tau_{\rm ren} {\vec n}(X) \times] {\dv n}(X) \notag \\
  & - \frac{1}{4 \pi^2 J} \left(\frac{2 m J}{\hbar^2}\right)^{1/2} {\cal F}_{i0} [\tau_{\beta} - \tau_{\rm STT} {\vec n}({\vec X}) \times] \partial_{X^i} {\vec n}({\vec X}), \label{eq:torque1a} \\
  \langle {\vec \sigma} \rangle_0(X)
  \equiv & \frac{1}{2 \pi^2} \left(\frac{2 m}{\hbar^2}\right)^{3/2} {\vec n}(X) \int {\rm d} \xi f(\xi) [-\Im g_{03}^{\rm R}(\xi)], \label{eq:torque1b} \\
  \tau_{\alpha}
  \equiv & J^{-1/2} \int {\rm d} \xi [-f^{\prime}(\xi)] [i g_{D 1}^{< (1)}(\xi)/2], \label{eq:torque1c} \\
  \tau_{\rm ren}
  \equiv & J^{-1/2} \int {\rm d} \xi \{f(\xi) [-\Im g_{D 2}^{\rm R}(\xi)] - [-f^{\prime}(\xi)] [i g_{D 2}^{< (1)}(\xi)/2]\}, \label{eq:torque1d} \\
  \tau_{\beta}
  \equiv & -J^{1/2} \int {\rm d} \xi \{f(\xi) [-\Im g_{D \ast {\cal F} 1}^{\rm R}(\xi)] + [-f^{\prime}(\xi)] [g_{D \ast {\cal F} 1}^{< (1)}(\xi)/2 i]\}, \label{eq:torque1e} \\
  \tau_{\rm STT}
  \equiv & J^{1/2} \int {\rm d} \xi [-f^{\prime}(\xi)] [g_{D \ast {\cal F} 2}^{< (1)}(\xi)/2 i], \label{eq:torque1f}
\end{align} \label{eq:torque1}\end{subequations}
\end{widetext}
and spin torques by
\begin{align}
  {\vec \tau}(X)
  = & {\vec n}(X) \times J \langle {\vec \sigma} \rangle(X) \notag \\
  = & -\frac{\hbar}{4 \pi^2} \left(\frac{2 m J}{\hbar^2}\right)^{3/2} [\tau_{\rm ren} + \tau_{\alpha} {\vec n}(X) \times] {\dv n}(X) \notag \\
  & - \frac{1}{4 \pi^2} \left(\frac{2 m J}{\hbar^2}\right)^{1/2} \notag \\
  & \times {\cal F}_{i0} [\tau_{\rm STT} + \tau_{\beta} {\vec n}({\vec X}) \times] \partial_{X^i} {\vec n}({\vec X}). \label{eq:torque2}
\end{align}
$\tau_{\rm ren}$ and $\tau_{\alpha}$ are the dimensionless spin renormalization and Gilbert damping,
while $\tau_{\rm STT}$ and $\tau_{\beta}$ are the STT and $\beta$-term.
To evaluate Eqs.~\eqref{eq:torque1c}-\eqref{eq:torque1f},
we carry out numerical integrals by putting the energy unit $J = 1$, the momentum cutoff $\Lambda = 10^3$, and temperature $T = 10^{-3}$
and dividing the energy interval $|\xi| < 5$ into $2^{17}$ subintervals.
In Fig.~\ref{fig:torque}, we show their chemical-potential dependences for different $\gamma_{\rm i}$ and $\gamma_{\rm s}$.
We also show the previous results obtained by the small-amplitude~\cite{JPSJ.75.113706} and the SU($2$) gauge transformation formalisms~\cite{JPSJ.76.063710} at zero temperature,
\begin{subequations} \begin{align}
  \tau_{\rm ren}
  = & (\mu_+^{3/2} - \mu_-^{3/2})/3 J^{3/2}, \label{eq:torque3a} \\
  \tau_{\alpha}
  = & \gamma_{\rm s} (\mu_+^{1/2} + \mu_-^{1/2})^2/3 J^{3/2}, \label{eq:torque3b} \\
  \tau_{\rm STT}
  = & \frac{1}{3 J^{1/2}} \sum_{\sigma} \frac{\sigma \mu_{\sigma}^{3/2}}{\gamma_{\rm i} \mu_{\sigma}^{1/2} + \gamma_{\rm s} (2 \mu_{-\sigma}^{1/2} + \mu_{\sigma}^{1/2})/3} \notag \\
  & \to 2 J^{1/2} (3 \gamma_{\rm i} + 5 \gamma_{\rm s})/9 (\gamma_{\rm i} + \gamma_{\rm s})^2
  \equiv \tau_{{\rm STT} \infty}, \label{eq:torque3c} \\
  \tau_{\beta}
  = & 2 \gamma_{\rm s} (\mu_+^{1/2} + \mu_-^{1/2}) \tau_{\rm STT}/3 J, \label{eq:torque3d}
\end{align} \label{eq:torque3}\end{subequations}
with $\mu_{\sigma} \equiv \mu + \sigma J$.
Note that $\tau_{{\rm STT} \infty}$ in Eq.~\eqref{eq:torque3c} is $\tau_{\rm STT}$ for the $\mu \to \infty$ limit.
Our results completely coincide with these previous ones.

\section{Discussion and summary} \label{sec:discussion}
Let us clarify how the SU($2$) gauge transformation formalism corresponds to the gradient expansion formalism.
In the former, the Green's function is diagonalized by a unitary matrix $U \equiv {\vec u} \cdot {\vec \sigma}$
with ${\vec u} \equiv [\sin \theta/2 \cos \phi, \sin \theta/2 \sin \phi, \cos \theta/2]$,
which transforms the magnetization ${\vec n} \equiv [\sin \theta \cos \phi, \sin \theta \sin \phi, \cos \theta]$ to ${\vec z}$
and yields the SU($2$) gauge field
$A_{\mu} \equiv -i U^{\dag} \partial_{X^{\mu}} U = {\vec u} \times \partial_{X^{\mu}} {\vec u} \cdot {\vec \sigma}$~\cite{PhysRevLett.92.086601,JPSJ.76.063710,PhysRevB.77.214429}.
Thus, spacetime gradients of the magnetization are described by this SU($2$) gauge field.
In the latter, $\partial_{X^{\mu}} (G_0^{\rm R})^{-1} \propto \partial_{X^{\mu}} {\vec n} \cdot {\vec \sigma}$ in Eqs.~\eqref{eq:first2} and \eqref{eq:second2}
is transformed to $U^{\dag} \partial_{X^{\mu}} (G_0^{\rm R})^{-1} U \propto 2 {\vec z} \times {\vec A}_{\mu} \cdot {\vec \sigma}$ and thus plays the same role as the SU($2$) gauge field.

We emphasize that there is another source of the SU($2$) gauge field in the SU($2$) gauge transformation formalism~\cite{JPSJ.76.063710}.
Magnetic impurities, which are quenched in the original frame, become dynamical and nonuniform in the adiabatic frame and yield the SU($2$) gauge field.
If this contribution is not taken into account, the Gilbert damping vanishes, and the $\beta$-term is not fully reproduced when magnetic impurities are anisotropic.
In the gradient expansion, we do not rely on the SU($2$) gauge transformation and hence do not encounter such difficulty.

In summary, we demonstrated how the gradient expansion works for calculating spin torques quantum-mechanically.
We derived the gradient expansion up to the fourth order with respect to the relative coordinates with abelian or nonabelian gauge fields being taken into account,
which enables us to investigate nonlinear responses with respect to spacetime gradients as well as field strengths.
We applied this formalism to a $3$d ferromagnetic metal with nonmagnetic and magnetic impurities
and successfully reproduced the previous results on the spin renormalization, Gilbert damping, STT, and $\beta$-term.
The greatest advantage of our formalism is that we do not assume any assumptions such as small transverse fluctuations
or suffer from the SU($2$) gauge field arising from magnetic impurities or SOIs.
Our central results Eqs.~\eqref{eq:first1}-\eqref{eq:spin} serve as first-principles formulas of spin torques.

\begin{acknowledgments}
  We thank J.~Fujimoto and G.~Tatara for discussion and reading our manuscript.
  This work was supported by RIKEN Special Postdoctoral Researcher Program.
\end{acknowledgments}
\appendix
\begin{widetext}
\section{Derivation of Eq.~\eqref{eq:moyal2} and more} \label{app:derivation}
In this Appendix, we evaluate the Wigner representation of convolution Eq.~\eqref{eq:moyal1},
\begin{align}
  \widetilde{A \ast B}(X_{12}, p_{12})
  = & \int {\rm d}^D x_{12} \int {\rm d}^D x_3 e^{p_{12 \mu} x_{12}^{\mu}/i \hbar} W(X_{12}, x_1) {\hat A}(x_1, x_3) {\hat B}(x_3, x_2) W(x_2, X_{12}) \notag \\
  = & \int {\rm d}^D x_{12} \int {\rm d}^D x_3 \int \frac{{\rm d}^D p_{13}}{(2 \pi \hbar)^D} \int \frac{{\rm d}^D p_{32}}{(2 \pi \hbar)^D}
  e^{p_{12 \mu} x_{12}^{\mu}/i \hbar} e^{-p_{13 \mu} x_{13}^{\mu}/i \hbar} e^{-p_{32 \mu} x_{32}^{\mu}/i \hbar} \notag \\
  & \times W(X_{12}, x_1) W(x_1, X_{13}) {\tilde A}(X_{13}, p_{13}) W(X_{13}, x_3) \notag \\
  & \times W(x_3, X_{32}) {\tilde B}(X_{32}, p_{32}) W(X_{32}, x_2) W(x_2, X_{12}) \label{eq:wigner2a} \\
  = & \int {\rm d}^D x_{12} \int {\rm d}^D x_3 \int \frac{{\rm d}^D p_{13}}{(2 \pi \hbar)^D} \int \frac{{\rm d}^D p_{32}}{(2 \pi \hbar)^D}
  e^{p_{12 \mu} x_{12}^{\mu}/i \hbar} e^{-p_{13 \mu} x_{13}^{\mu}/i \hbar} e^{-p_{32 \mu} x_{32}^{\mu}/i \hbar} \notag \\
  & \times [W(X_{12}, x_1) W(x_1, X_{13}) W(X_{13}, X_{12})] [W(X_{12}, X_{13}) {\tilde A}(X_{13}, p_{13}) W(X_{13}, X_{12})] \notag \\
  & \times [W(X_{12}, X_{13}) W(X_{13}, x_3) W(x_3, X_{32}) W(X_{32}, X_{12})] [W(X_{12}, X_{32}) {\tilde B}(X_{32}, p_{32}) W(X_{32}, X_{12})] \notag \\
  & \times [W(X_{12}, X_{32}) W(X_{32}, x_2) W(x_2, X_{12})]. \label{eq:wigner2b}
\end{align}
Here we insert the identities $W(X_{13}, X_{12}) W(X_{12}, X_{13}) = W(X_{32}, X_{12}) W(X_{12}, X_{32}) = 1$,
which corresponds to transformation of the Wilson line as shown in Fig.~\ref{fig:wilson}.
\begin{figure*}
  \centering
  \includegraphics[clip,width=\textwidth]{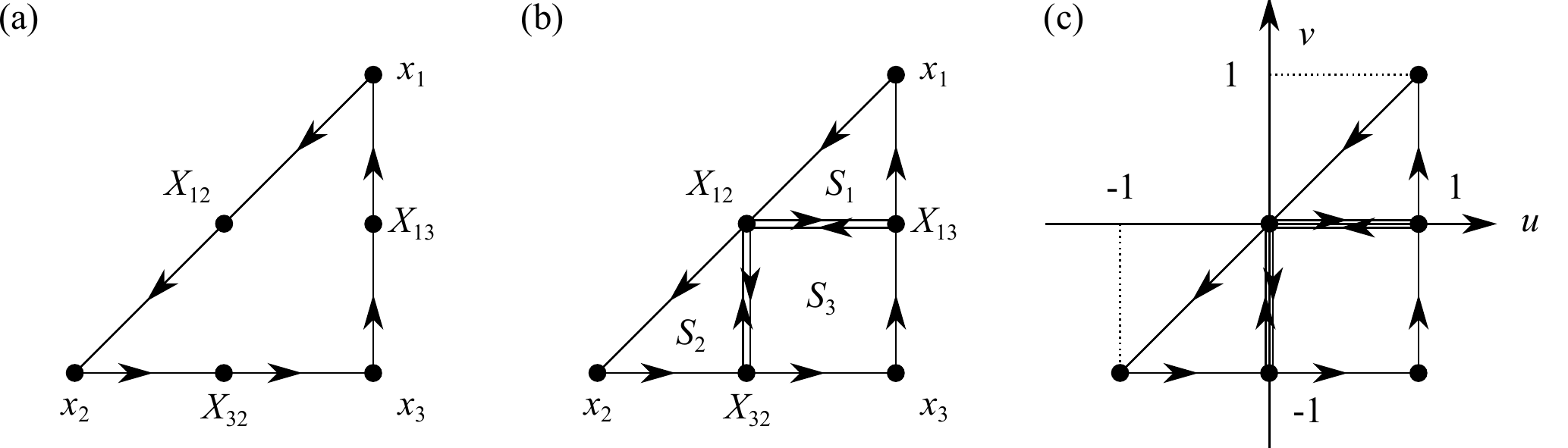}
  \caption{%
  Transformation from 
  (a) the Wilson line corresponding to Eq.~\eqref{eq:wigner2a} to
  (b) that to Eq.~\eqref{eq:wigner2b}.
  The plane spanned by $x_1, x_2, x_3$ is divided into three planes, $S_1$, $S_2$, and $S_3$.
  (c) Three planes are parametrized by $y = X_{12} + u x_{32}/2 + v x_{13}/2$.%
  } \label{fig:wilson}
\end{figure*}
We define three planes $S_1$, $S_2$, and $S_3$ as in Fig.~\ref{fig:wilson}(b), whose origin is $X_{12}$.
To evaluate the first factor of the integrand in Eq.~\eqref{eq:wigner2b}, we use the nonabelian Stokes theorem~\cite{Arefeva1980,PhysRevD.22.3090},
\begin{align}
  W(X_{12}, x_1) W(x_1, X_{13}) W(X_{13}, X_{12})
  = & S \exp \left[-\frac{1}{2 i \hbar} \int_{S_1} {\rm d} y^{\mu} {\rm d} y^{\nu} W(X_{12}, y) {\cal F}_{\mu \nu}(y) W(y, X_{12})\right] \notag \\
  = & S \exp \left[\frac{1}{8 i \hbar} \int_0^1 {\rm d} u \int_0^u {\rm d} v (x_{13}^{\mu} x_{32}^{\nu} - x_{13}^{\nu} x_{32}^{\mu})
  W(X_{12}, y) {\cal F}_{\mu \nu}(y) W(y, X_{12})\right], \label{eq:wigner3}
\end{align}
in which $S$ is the surface-ordered product~\cite{Arefeva1980,PhysRevD.22.3090}, and $y = X_{12} + u x_{32}/2 + v x_{13}/2$.
By using the Taylor expansion around $X_{12}$, we obtain
\begin{subequations} \begin{align}
  {\cal F}_{\mu \nu}(y)
  = & e^{(u x_{32}^{\lambda} + v x_{13}^{\lambda}) \partial_{X_{12}^{\lambda}}/2} {\cal F}_{\mu \nu} \notag \\
  = & {\cal F}_{\mu \nu} + (u x_{32}^{\lambda} + v x_{13}^{\lambda}) \partial_{X_{12}^{\lambda}} {\cal F}_{\mu \nu}/2 \notag \\
  & + (u x_{32}^{\lambda_1} + v x_{13}^{\lambda_1}) (u x_{32}^{\lambda_2} + v x_{13}^{\lambda_2}) \partial_{X_{12}^{\lambda_1}} \partial_{X_{12}^{\lambda_2}} {\cal F}_{\mu \nu}/8 + O(x^3), \label{eq:wigner4a} \\
  W(y, X_{12})
  = & 1 - \frac{1}{2 i \hbar} \int_0^1 {\rm d} w (u x_{32}^{\mu} + v x_{13}^{\mu}) [{\cal A}_{\mu} + w (u x_{32}^{\nu} + v x_{13}^{\nu}) \partial_{X_{12}^{\nu}} {\cal A}_{\mu}/2] \notag \\
  & + \frac{1}{4 (i \hbar)^2} \int_0^1 {\rm d} w_1 \int_0^{w_1} {\rm d} w_2 (u x_{32}^{\mu_1} + v x_{13}^{\mu_1}) (u x_{32}^{\mu_2} + v x_{13}^{\mu_2}) {\cal A}_{\mu_1} {\cal A}_{\mu_2} + O(x^3) \notag \\
  = & 1 - (u x_{32}^{\mu} + v x_{13}^{\mu}) {\cal A}_{\mu}/2 i \hbar - (u x_{32}^{\mu} + v x_{13}^{\mu}) (u x_{32}^{\nu} + v x_{13}^{\nu}) \partial_{X_{12}^{\nu}} {\cal A}_{\mu}/8 i \hbar \notag \\
  & + (u x_{32}^{\mu_1} + v x_{13}^{\mu_1}) (u x_{32}^{\mu_2} + v x_{13}^{\mu_2}) {\cal A}_{\mu_1} {\cal A}_{\mu_2}/8 (i \hbar)^2 + O(x^3), \label{eq:wigner4b} \\
  W(X_{12}, y) {\cal F}_{\mu \nu}(y) W(y, X_{12})
  = & {\cal F}_{\mu \nu}
  + (u x_{32}^{\lambda} + v x_{13}^{\lambda}) (\partial_{X_{12}^{\lambda}} {\cal F}_{\mu \nu} + [{\cal A}_{\lambda}, {\cal F}_{\mu \nu}]/i \hbar)/2 \notag \\
  & + (u x_{32}^{\lambda_1} + v x_{13}^{\lambda_1}) (u x_{32}^{\lambda_2} + v x_{13}^{\lambda_2})
  [\partial_{X_{12}^{\lambda_1}} \partial_{X_{12}^{\lambda_2}} {\cal F}_{\mu \nu} + [\partial_{X_{12}^{\lambda_2}} {\cal A}_{\lambda_1}, {\cal F}_{\mu \nu}]/i \hbar \notag \\
  & + 2 [{\cal A}_{\lambda_1}, \partial_{X_{12}^{\lambda_2}} {\cal F}_{\mu \nu}]/i \hbar + [{\cal A}_{\lambda_1}, [{\cal A}_{\lambda_2}, {\cal F}_{\mu \nu}]]/(i \hbar)^2]/8 + O(x^3) \notag \\
  = & {\cal F}_{\mu \nu} + (u x_{32}^{\lambda} + v x_{13}^{\lambda}) D_{X_{12}^{\lambda}} {\cal F}_{\mu \nu}/2 \notag \\
  & + (u x_{32}^{\lambda_1} + v x_{13}^{\lambda_1}) (u x_{32}^{\lambda_2} + v x_{13}^{\lambda_2}) D_{X_{12}^{\lambda_1}} D_{X_{12}^{\lambda_2}} {\cal F}_{\mu \nu}/8 + O(x^3) \notag \\
  = & e^{(u x_{32}^{\lambda} + v x_{13}^{\lambda}) D_{X_{12}^{\lambda}}/2} {\cal F}_{\mu \nu}. \label{eq:wigner4c}
\end{align} \label{eq:wigner4}\end{subequations}
Here and below we omit the argument $X_{12}$ for simplicity.
Then, we express Eq.~\eqref{eq:wigner3} as
\begin{align}
  W(X_{12}, x_1) W(x_1, X_{13}) W(X_{13}, X_{12})
  = & 1 + \frac{1}{8 i \hbar} \int_0^1 {\rm d} u \int_0^u {\rm d} v (x_{13}^{\mu} x_{32}^{\nu} - x_{13}^{\nu} x_{32}^{\mu})
  [{\cal F}_{\mu \nu} + (u x_{32}^{\lambda} + v x_{13}^{\lambda}) D_{X_{12}^{\lambda}} {\cal F}_{\mu \nu}/2 \notag \\
  & + (u x_{32}^{\lambda_1} + v x_{13}^{\lambda_1}) (u x_{32}^{\lambda_2} + v x_{13}^{\lambda_2}) D_{X_{12}^{\lambda_1}} D_{X_{12}^{\lambda_2}} {\cal F}_{\mu \nu}/8] \notag \\
  & + \frac{1}{64 (i \hbar)^2} \int_0^1 {\rm d} u_1 \int_0^{u_1} {\rm d} v_1 \int_0^{u_1} {\rm d} u_2 \int_0^{u_2} {\rm d} v_2 \notag \\
  & \times (x_{13}^{\mu_1} x_{32}^{\nu_1} - x_{13}^{\nu_1} x_{32}^{\mu_1}) (x_{13}^{\mu_2} x_{32}^{\nu_2} - x_{13}^{\nu_2} x_{32}^{\mu_2})
  {\cal F}_{\mu_1 \nu_1} {\cal F}_{\mu_2 \nu_2} + O(x^5) \notag \\
  = & 1 + (x_{13}^{\mu} x_{32}^{\nu} - x_{13}^{\nu} x_{32}^{\mu}) {\cal F}_{\mu \nu}/16 i \hbar
  + (x_{13}^{\mu} x_{32}^{\nu} - x_{13}^{\nu} x_{32}^{\mu}) (2 x_{32}^{\lambda} + x_{13}^{\lambda}) D_{X_{12}^{\lambda}} {\cal F}_{\mu \nu}/96 i \hbar \notag \\
  & + (x_{13}^{\mu_1} x_{32}^{\nu_1} - x_{13}^{\nu_1} x_{32}^{\mu_1}) (x_{13}^{\mu_2} x_{32}^{\nu_2} - x_{13}^{\nu_2} x_{32}^{\mu_2}) {\cal F}_{\mu_1 \nu_1} {\cal F}_{\mu_2 \nu_2}/512 (i \hbar)^2 \notag \\
  & + (x_{13}^{\mu} x_{32}^{\nu} - x_{13}^{\nu} x_{32}^{\mu})
  [6 x_{32}^{\lambda_1} x_{32}^{\lambda_2} + 3 (x_{32}^{\lambda_1} x_{13}^{\lambda_2} + x_{32}^{\lambda_2} x_{13}^{\lambda_1}) + 2 x_{13}^{\lambda_1} x_{13}^{\lambda_2}] \notag \\
  & \times D_{X_{12}^{\lambda_1}} D_{X_{12}^{\lambda_2}} {\cal F}_{\mu \nu}/1536 i \hbar + O(x^5). \label{eq:wigner5}
\end{align}
The other factors in Eq.~\eqref{eq:wigner2b} are similarly obtained as
\begin{subequations} \begin{align}
  W(X_{12}, X_{13}) {\tilde A}(X_{13}, p_{13}) W(X_{13}, X_{12})
  = & e^{x_{32}^{\lambda} D_{X_{12}^{\lambda}}/2} {\tilde A}(p_{13}), \label{eq:wigner6a} \\
  W(X_{12}, X_{32}) {\tilde B}(X_{32}, p_{32}) W(X_{32}, X_{12})
  = & e^{-x_{13}^{\lambda} D_{X_{12}^{\lambda}}/2} {\tilde B}(p_{32}), \label{eq:wigner6b} \\
  W(X_{12}, X_{13}) W(X_{13}, x_3) W(x_3, X_{32}) W(X_{32}, X_{12})
  = & 1 + (x_{13}^{\mu} x_{32}^{\nu} - x_{13}^{\nu} x_{32}^{\mu}) {\cal F}_{\mu \nu}/8 i \hbar \notag \\
  & + (x_{13}^{\mu} x_{32}^{\nu} - x_{13}^{\nu} x_{32}^{\mu}) (x_{32}^{\lambda} - x_{13}^{\lambda}) D_{X_{12}^{\lambda}} {\cal F}_{\mu \nu}/32 i \hbar \notag \\
  & + (x_{13}^{\mu_1} x_{32}^{\nu_1} - x_{13}^{\nu_1} x_{32}^{\mu_1}) (x_{13}^{\mu_2} x_{32}^{\nu_2} - x_{13}^{\nu_2} x_{32}^{\mu_2}) {\cal F}_{\mu_1 \nu_1} {\cal F}_{\mu_2 \nu_2}/128 (i \hbar)^2 \notag \\
  & + (x_{13}^{\mu} x_{32}^{\nu} - x_{13}^{\nu} x_{32}^{\mu})
  [4 x_{32}^{\lambda_1} x_{32}^{\lambda_2} - 3 (x_{32}^{\lambda_1} x_{13}^{\lambda_2} + x_{32}^{\lambda_2} x_{13}^{\lambda_1}) + 4 x_{13}^{\lambda_1} x_{13}^{\lambda_2}] \notag \\
  & \times D_{X_{12}^{\lambda_1}} D_{X_{12}^{\lambda_2}} {\cal F}_{\mu \nu}/768 i \hbar + O(x^5), \label{eq:wigner6c} \\
  W(X_{12}, X_{32}) W(X_{32}, x_2) W(x_2, X_{12})
  = & 1 + (x_{13}^{\mu} x_{32}^{\nu} - x_{13}^{\nu} x_{32}^{\mu}) {\cal F}_{\mu \nu}/16 i \hbar \notag \\
  & - (x_{13}^{\mu} x_{32}^{\nu} - x_{13}^{\nu} x_{32}^{\mu}) (x_{32}^{\lambda} + 2 x_{13}^{\lambda}) D_{X_{12}^{\lambda}} {\cal F}_{\mu \nu}/96 i \hbar \notag \\
  & + (x_{13}^{\mu_1} x_{32}^{\nu_1} - x_{13}^{\nu_1} x_{32}^{\mu_1}) (x_{13}^{\mu_2} x_{32}^{\nu_2} - x_{13}^{\nu_2} x_{32}^{\mu_2}) {\cal F}_{\mu_1 \nu_1} {\cal F}_{\mu_2 \nu_2}/512 (i \hbar)^2 \notag \\
  & + (x_{13}^{\mu} x_{32}^{\nu} - x_{13}^{\nu} x_{32}^{\mu})
  [2 x_{32}^{\lambda_1} x_{32}^{\lambda_2} + 3 (x_{32}^{\lambda_1} x_{13}^{\lambda_2} + x_{32}^{\lambda_2} x_{13}^{\lambda_1}) + 6 x_{13}^{\lambda_1} x_{13}^{\lambda_2}] \notag \\
  & \times D_{X_{12}^{\lambda_1}} D_{X_{12}^{\lambda_2}} {\cal F}_{\mu \nu}/1536 i \hbar + O(x^5). \label{eq:wigner6d}
\end{align} \label{eq:wigner6}\end{subequations}
Thus, the integrand in Eq.~\eqref{eq:wigner2b} is given by
\begin{subequations} \begin{align}
  O(1)
  = & {\tilde A}(p_{13}) {\tilde B}(p_{32}), \label{eq:wigner7a} \\
  O(x)
  = & [x_{32}^{\lambda} D_{X_{12}^{\lambda}} {\tilde A}(p_{13}) {\tilde B}(p_{32}) - x_{13}^{\lambda} {\tilde A}(p_{13}) D_{X_{12}^{\lambda}} {\tilde B}(p_{32})]/2, \label{eq:wigner7b} \\
  O(x^2)
  = & (x_{13}^{\mu} x_{32}^{\nu} - x_{13}^{\nu} x_{32}^{\mu}) 
  [{\cal F}_{\mu \nu} {\tilde A}(p_{13}) {\tilde B}(p_{32}) + 2 {\tilde A}(p_{13}) {\cal F}_{\mu \nu} {\tilde B}(p_{32}) + {\tilde A}(p_{13}) {\tilde B}(p_{32}) {\cal F}_{\mu \nu}]/16 i \hbar \label{eq:wigner7c} \\
  & + [x_{32}^{\lambda_1} x_{32}^{\lambda_2} D_{X_{12}^{\lambda_1}} D_{X_{12}^{\lambda_2}} {\tilde A}(p_{13}) {\tilde B}(p_{32})
  - 2 x_{32}^{\lambda_1} x_{13}^{\lambda_2} D_{X_{12}^{\lambda_1}} {\tilde A}(p_{13}) D_{X_{12}^{\lambda_2}} {\tilde B}(p_{32}) \notag \\
  & + x_{13}^{\lambda_1} x_{13}^{\lambda_2} {\tilde A}(p_{13}) D_{X_{12}^{\lambda_1}} D_{X_{12}^{\lambda_2}} {\tilde B}(p_{32})]/8, \label{eq:wigner7d} \\
  O(x^3)
  = & (x_{13}^{\mu} x_{32}^{\nu} - x_{13}^{\nu} x_{32}^{\mu}) 
  \{{\cal F}_{\mu \nu} [x_{32}^{\lambda} D_{X_{12}^{\lambda}} {\tilde A}(p_{13}) {\tilde B}(p_{32}) - x_{13}^{\lambda} {\tilde A}(p_{13}) D_{X_{12}^{\lambda}} {\tilde B}(p_{32})] \notag \\
  & + 2 [x_{32}^{\lambda} D_{X_{12}^{\lambda}} {\tilde A}(p_{13}) {\cal F}_{\mu \nu} {\tilde B}(p_{32}) - x_{13}^{\lambda} {\tilde A}(p_{13}) {\cal F}_{\mu \nu} D_{X_{12}^{\lambda}} {\tilde B}(p_{32})] \notag \\
  & + [x_{32}^{\lambda} D_{X_{12}^{\lambda}} {\tilde A}(p_{13}) {\tilde B}(p_{32}) - x_{13}^{\lambda} {\tilde A}(p_{13}) D_{X_{12}^{\lambda}} {\tilde B}(p_{32})] {\cal F}_{\mu \nu}\}/32 i \hbar \label{eq:wigner7e} \\
  & + (x_{13}^{\mu} x_{32}^{\nu} - x_{13}^{\nu} x_{32}^{\mu})
  [(2 x_{32}^{\lambda} + x_{13}^{\lambda}) D_{X_{12}^{\lambda}} {\cal F}_{\mu \nu} {\tilde A}(p_{13}) {\tilde B}(p_{32})
  + 3 (x_{32}^{\lambda} - x_{13}^{\lambda}) {\tilde A}(p_{13}) D_{X_{12}^{\lambda}} {\cal F}_{\mu \nu} {\tilde B}(p_{32}) \notag \\
  & - (x_{32}^{\lambda} + 2 x_{13}^{\lambda}) {\tilde A}(p_{13}) {\tilde B}(p_{32}) D_{X_{12}^{\lambda}} {\cal F}_{\mu \nu}]/96 i \hbar \label{eq:wigner7f} \\
  & + [x_{32}^{\lambda_1} x_{32}^{\lambda_2} x_{32}^{\lambda_3} D_{X_{12}^{\lambda_1}} D_{X_{12}^{\lambda_2}} D_{X_{12}^{\lambda_3}} {\tilde A}(p_{13}) {\tilde B}(p_{32})
  - 3 x_{32}^{\lambda_1} x_{32}^{\lambda_2} x_{13}^{\lambda_3} D_{X_{12}^{\lambda_1}} D_{X_{12}^{\lambda_2}} {\tilde A}(p_{13}) D_{X_{12}^{\lambda_3}} {\tilde B}(p_{32}) \notag \\
  & + 3 x_{32}^{\lambda_1} x_{13}^{\lambda_2} x_{13}^{\lambda_3} D_{X_{12}^{\lambda_1}} {\tilde A}(p_{13}) D_{X_{12}^{\lambda_2}} D_{X_{12}^{\lambda_3}} {\tilde B}(p_{32})
  - x_{13}^{\lambda_1} x_{13}^{\lambda_2} x_{13}^{\lambda_3} {\tilde A}(p_{13}) D_{X_{12}^{\lambda_1}} D_{X_{12}^{\lambda_2}} D_{X_{12}^{\lambda_3}} {\tilde B}(p_{32})]/48, \label{eq:wigner7g} \\
  O(x^4)
  = & (x_{13}^{\mu_1} x_{32}^{\nu_1} - x_{13}^{\nu_1} x_{32}^{\mu_1}) (x_{13}^{\mu_2} x_{32}^{\nu_2} - x_{13}^{\nu_2} x_{32}^{\mu_2}) \notag \\
  & \times [{\cal F}_{\mu_1 \nu_1} {\cal F}_{\mu_2 \nu_2} {\tilde A}(p_{13}) {\tilde B}(p_{32}) + 4 {\tilde A}(p_{13}) {\cal F}_{\mu_1 \nu_1} {\cal F}_{\mu_2 \nu_2} {\tilde B}(p_{32})
  + {\tilde A}(p_{13}) {\tilde B}(p_{32}) {\cal F}_{\mu_1 \nu_1} {\cal F}_{\mu_2 \nu_2} \notag \\
  & + 4 {\cal F}_{\mu_1 \nu_1} {\tilde A}(p_{13}) {\cal F}_{\mu_2 \nu_2} {\tilde B}(p_{32}) + 4 {\tilde A}(p_{13}) {\cal F}_{\mu_1 \nu_1} {\tilde B}(p_{32}) {\cal F}_{\mu_2 \nu_2}
  + 2 {\cal F}_{\mu_1 \nu_1} {\tilde A}(p_{13}) {\tilde B}(p_{32}) {\cal F}_{\mu_2 \nu_2}]/512 (i \hbar)^2 \label{eq:wigner7h} \\
  & + (x_{13}^{\mu} x_{32}^{\nu} - x_{13}^{\nu} x_{32}^{\mu})
  \{{\cal F}_{\mu \nu} [x_{32}^{\lambda_1} x_{32}^{\lambda_2} D_{X_{12}^{\lambda_1}} D_{X_{12}^{\lambda_2}} {\tilde A}(p_{13}) {\tilde B}(p_{32})
  - 2 x_{32}^{\lambda_1} x_{13}^{\lambda_2} D_{X_{12}^{\lambda_1}} {\tilde A}(p_{13}) D_{X_{12}^{\lambda_2}} {\tilde B}(p_{32}) \notag \\
  & + x_{13}^{\lambda_1} x_{13}^{\lambda_2} {\tilde A}(p_{13}) D_{X_{12}^{\lambda_1}} D_{X_{12}^{\lambda_2}} {\tilde B}(p_{32})]
  + 2 [x_{32}^{\lambda_1} x_{32}^{\lambda_2} D_{X_{12}^{\lambda_1}} D_{X_{12}^{\lambda_2}} {\tilde A}(p_{13}) {\cal F}_{\mu \nu} {\tilde B}(p_{32}) \notag \\
  & - 2 x_{32}^{\lambda_1} x_{13}^{\lambda_2} D_{X_{12}^{\lambda_1}} {\tilde A}(p_{13}) {\cal F}_{\mu \nu} D_{X_{12}^{\lambda_2}} {\tilde B}(p_{32})
  + x_{13}^{\lambda_1} x_{13}^{\lambda_2} {\tilde A}(p_{13}) {\cal F}_{\mu \nu} D_{X_{12}^{\lambda_1}} D_{X_{12}^{\lambda_2}} {\tilde B}(p_{32})] \notag \\
  & + [x_{32}^{\lambda_1} x_{32}^{\lambda_2} D_{X_{12}^{\lambda_1}} D_{X_{12}^{\lambda_2}} {\tilde A}(p_{13}) {\tilde B}(p_{32})
  - 2 x_{32}^{\lambda_1} x_{13}^{\lambda_2} D_{X_{12}^{\lambda_1}} {\tilde A}(p_{13}) D_{X_{12}^{\lambda_2}} {\tilde B}(p_{32}) \notag \\
  & + x_{13}^{\lambda_1} x_{13}^{\lambda_2} {\tilde A}(p_{13}) D_{X_{12}^{\lambda_1}} D_{X_{12}^{\lambda_2}} {\tilde B}(p_{32})] {\cal F}_{\mu \nu}\}/128 i \hbar \label{eq:wigner7i} \\
  & + (x_{13}^{\mu} x_{32}^{\nu} - x_{13}^{\nu} x_{32}^{\mu})
  \{(2 x_{32}^{\lambda_1} + x_{13}^{\lambda_1}) D_{X_{12}^{\lambda_1}} {\cal F}_{\mu \nu}
  [x_{32}^{\lambda_2} D_{X_{12}^{\lambda_2}} {\tilde A}(p_{13}) {\tilde B}(p_{32}) - x_{13}^{\lambda_2} {\tilde A}(p_{13}) D_{X_{12}^{\lambda_2}} {\tilde B}(p_{32})] \notag \\
  & + 3 (x_{32}^{\lambda_1} - x_{13}^{\lambda_1}) [x_{32}^{\lambda_2} D_{X_{12}^{\lambda_2}} {\tilde A}(p_{13}) D_{X_{12}^{\lambda_1}} {\cal F}_{\mu \nu} {\tilde B}(p_{32})
  - x_{13}^{\lambda_2} {\tilde A}(p_{13}) D_{X_{12}^{\lambda_1}} {\cal F}_{\mu \nu} D_{X_{12}^{\lambda_2}} {\tilde B}(p_{32})] \notag \\
  & - (x_{32}^{\lambda_1} + 2 x_{13}^{\lambda_1}) [x_{32}^{\lambda_2} D_{X_{12}^{\lambda_2}} {\tilde A}(p_{13}) {\tilde B}(p_{32}) - x_{13}^{\lambda_2} {\tilde A}(p_{13}) D_{X_{12}^{\lambda_2}} {\tilde B}(p_{32})]
  D_{X_{12}^{\lambda_1}} {\cal F}_{\mu \nu}\}/96 i \hbar \label{eq:wigner7j} \\
  & + (x_{13}^{\mu} x_{32}^{\nu} - x_{13}^{\nu} x_{32}^{\mu})
  \{[6 x_{32}^{\lambda_1} x_{32}^{\lambda_2} + 3 (x_{32}^{\lambda_1} x_{13}^{\lambda_2} + x_{32}^{\lambda_2} x_{13}^{\lambda_1}) + 2 x_{13}^{\lambda_1} x_{13}^{\lambda_2}]
  D_{X_{12}^{\lambda_1}} D_{X_{12}^{\lambda_2}} {\cal F}_{\mu \nu} {\tilde A}(p_{13}) {\tilde B}(p_{32}) \notag \\
  & + [8 x_{32}^{\lambda_1} x_{32}^{\lambda_2} - 6 (x_{32}^{\lambda_1} x_{13}^{\lambda_2} + x_{32}^{\lambda_2} x_{13}^{\lambda_1}) + 8 x_{13}^{\lambda_1} x_{13}^{\lambda_2}]
  {\tilde A}(p_{13}) D_{X_{12}^{\lambda_1}} D_{X_{12}^{\lambda_2}} {\cal F}_{\mu \nu} {\tilde B}(p_{32}) \notag \\
  & + [2 x_{32}^{\lambda_1} x_{32}^{\lambda_2} + 3 (x_{32}^{\lambda_1} x_{13}^{\lambda_2} + x_{32}^{\lambda_2} x_{13}^{\lambda_1}) + 6 x_{13}^{\lambda_1} x_{13}^{\lambda_2}]
  {\tilde A}(p_{13}) {\tilde B}(p_{32}) D_{X_{12}^{\lambda_1}} D_{X_{12}^{\lambda_2}} {\cal F}_{\mu \nu}\}/1536 i \hbar \label{eq:wigner7k} \\
  & + [x_{32}^{\lambda_1} x_{32}^{\lambda_2} x_{32}^{\lambda_3} x_{32}^{\lambda_4} D_{X_{12}^{\lambda_1}} D_{X_{12}^{\lambda_2}} D_{X_{12}^{\lambda_3}} D_{X_{12}^{\lambda_4}} {\tilde A}(p_{13}) {\tilde B}(p_{32})
  - 4 x_{32}^{\lambda_1} x_{32}^{\lambda_2} x_{32}^{\lambda_3} x_{13}^{\lambda_4} D_{X_{12}^{\lambda_1}} D_{X_{12}^{\lambda_2}} D_{X_{12}^{\lambda_3}} {\tilde A}(p_{13}) D_{X_{12}^{\lambda_4}} {\tilde B}(p_{32}) \notag \\
  & + 6 x_{32}^{\lambda_1} x_{32}^{\lambda_2} x_{13}^{\lambda_3} x_{13}^{\lambda_4} D_{X_{12}^{\lambda_1}} D_{X_{12}^{\lambda_2}} {\tilde A}(p_{13}) D_{X_{12}^{\lambda_3}} D_{X_{12}^{\lambda_4}} {\tilde B}(p_{32})
  - 4 x_{32}^{\lambda_1} x_{13}^{\lambda_2} x_{13}^{\lambda_3} x_{13}^{\lambda_4} D_{X_{12}^{\lambda_1}} {\tilde A}(p_{13}) D_{X_{12}^{\lambda_2}} D_{X_{12}^{\lambda_3}} D_{X_{12}^{\lambda_4}} {\tilde B}(p_{32}) \notag \\
  & + x_{13}^{\lambda_1} x_{13}^{\lambda_2} x_{13}^{\lambda_3} x_{13}^{\lambda_4} {\tilde A}(p_{13}) D_{X_{12}^{\lambda_1}} D_{X_{12}^{\lambda_2}} D_{X_{12}^{\lambda_3}} D_{X_{12}^{\lambda_4}} {\tilde B}(p_{32})]/384. \label{eq:wigner7l}
\end{align} \label{eq:wigner7}\end{subequations}
Now we can carry out the integrals by putting $p_{13} = p_{12} + q_{13}, p_{32} = p_{12} + q_{32}$.
Since $q_{13}$ and $q_{32}$ appear in the forms of ${\tilde A}(p_{12} + q_{13})$ and ${\tilde B}(p_{12} + q_{32})$,
$x_{13}^{\mu}$ and $x_{32}^{\mu}$ can be replaced with $i \hbar \partial_{p_{12 \mu}}$ acting on ${\tilde A}$ and ${\tilde B}$, respectively.
Finally, we obtain
\begin{subequations} \begin{align}
  {\tilde A} \star {\tilde B}
  = & {\tilde A} {\tilde B} + (i \hbar/2) {\cal P}_D({\tilde A}, {\tilde B}) + (i \hbar/2) {\cal P}_{\cal F}({\tilde A}, {\tilde B})
  + (1/2!) (i \hbar/2)^2 {\cal P}_{D^2}({\tilde A}, {\tilde B}) \notag \\
  & + (i \hbar/2)^2 {\cal P}_{D \ast {\cal F}}({\tilde A}, {\tilde B}) + (1/3) (i \hbar/2)^2 {\cal P}_{D {\cal F}}({\tilde A}, {\tilde B})
  + (1/3!) (i \hbar/2)^3 {\cal P}_{D^3}({\tilde A}, {\tilde B}) \notag \\
  & + (1/2!) (i \hbar/2)^2 {\cal P}_{{\cal F}^2}({\tilde A}, {\tilde B}) + (1/2!) (i \hbar/2)^3 {\cal P}_{D^2 \ast {\cal F}}({\tilde A}, {\tilde B})
  + (1/3) (i \hbar/2)^3 {\cal P}_{D \ast D {\cal F}}({\tilde A}, {\tilde B}) \notag \\
  & + (1/2!) (1/3) (i \hbar/2)^3 {\cal P}_{D^2 {\cal F}}({\tilde A}, {\tilde B}) + (1/4!) (i \hbar/2)^4 {\cal P}_{D^4}({\tilde A}, {\tilde B}), \label{eq:wigner8a} \\
  {\cal P}_D({\tilde A}, {\tilde B})
  \equiv & D_{X^{\lambda}} {\tilde A} \partial_{p_{\lambda}} {\tilde B} - \partial_{p_{\lambda}} {\tilde A} D_{X^{\lambda}} {\tilde B}, \label{eq:wigner8b} \\
  {\cal P}_{\cal F}({\tilde A}, {\tilde B})
  \equiv & ({\cal F}_{\mu \nu} \partial_{p_{\mu}} {\tilde A} \partial_{p_{\nu}} {\tilde B}
  + 2 \partial_{p_{\mu}} {\tilde A} {\cal F}_{\mu \nu} \partial_{p_{\nu}} {\tilde B}
  + \partial_{p_{\mu}} {\tilde A} \partial_{p_{\nu}} {\tilde B} {\cal F}_{\mu \nu})/4, \label{eq:wigner8c} \\
  {\cal P}_{D^2}({\tilde A}, {\tilde B})
  \equiv & D_{X^{\lambda_1}} D_{X^{\lambda_2}} {\tilde A} \partial_{p_{\lambda_1}} \partial_{p_{\lambda_2}} {\tilde B}
  - 2 D_{X^{\lambda_1}} \partial_{p_{\lambda_2}} {\tilde A} \partial_{p_{\lambda_1}} D_{X^{\lambda_2}} {\tilde B}
  + \partial_{p_{\lambda_1}} \partial_{p_{\lambda_2}} {\tilde A} D_{X^{\lambda_1}} D_{X^{\lambda_2}} {\tilde B}, \label{eq:wigner8d} \\
  {\cal P}_{D \ast {\cal F}}({\tilde A}, {\tilde B})
  \equiv & [{\cal F}_{\mu \nu} (D_{X^{\lambda}} \partial_{p_{\mu}} {\tilde A} \partial_{p_{\lambda}} \partial_{p_{\nu}} {\tilde B}
  - \partial_{p_{\lambda}} \partial_{p_{\mu}} {\tilde A} D_{X^{\lambda}} \partial_{p_{\nu}} {\tilde B}) \notag \\
  & + 2 (D_{X^{\lambda}} \partial_{p_{\mu}} {\tilde A} {\cal F}_{\mu \nu} \partial_{p_{\lambda}} \partial_{p_{\nu}} {\tilde B}
  - \partial_{p_{\lambda}} \partial_{p_{\mu}} {\tilde A} {\cal F}_{\mu \nu} D_{X^{\lambda}} \partial_{p_{\nu}} {\tilde B}) \notag \\
  & + (D_{X^{\lambda}} \partial_{p_{\mu}} {\tilde A} \partial_{p_{\lambda}} \partial_{p_{\nu}} {\tilde B}
  - \partial_{p_{\lambda}} \partial_{p_{\mu}} {\tilde A} D_{X^{\lambda}} \partial_{p_{\nu}} {\tilde B}) {\cal F}_{\mu \nu}]/4, \label{eq:wigner8e} \\
  {\cal P}_{D {\cal F}}({\tilde A}, {\tilde B})
  \equiv & [D_{X^{\lambda}} {\cal F}_{\mu \nu} (2 \partial_{p_{\mu}} {\tilde A} \partial_{p_{\lambda}} \partial_{p_{\nu}} {\tilde B}
  + \partial_{p_{\lambda}} \partial_{p_{\mu}} {\tilde A} \partial_{p_{\nu}} {\tilde B}) \notag \\
  & + 3 (\partial_{p_{\mu}} {\tilde A} D_{X^{\lambda}} {\cal F}_{\mu \nu} \partial_{p_{\lambda}} \partial_{p_{\nu}} {\tilde B}
  - \partial_{p_{\lambda}} \partial_{p_{\mu}} {\tilde A} D_{X^{\lambda}} {\cal F}_{\mu \nu} \partial_{p_{\nu}} {\tilde B}) \notag \\
  & - (\partial_{p_{\mu}} {\tilde A} \partial_{p_{\lambda}} \partial_{p_{\nu}} {\tilde B}
  + 2 \partial_{p_{\lambda}} \partial_{p_{\mu}} {\tilde A} \partial_{p_{\nu}} {\tilde B}) D_{X^{\lambda}} {\cal F}_{\mu \nu}]/4,  \label{eq:wigner8f} \\
  {\cal P}_{D^3}({\tilde A}, {\tilde B})
  \equiv & D_{X^{\lambda_1}} D_{X^{\lambda_2}} D_{X^{\lambda_3}} {\tilde A} \partial_{p_{\lambda_1}} \partial_{p_{\lambda_2}} \partial_{p_{\lambda_3}} {\tilde B}
  - 3 D_{X^{\lambda_1}} D_{X^{\lambda_2}} \partial_{p_{\lambda_3}} {\tilde A} \partial_{p_{\lambda_1}} \partial_{p_{\lambda_2}} D_{X^{\lambda_3}} {\tilde B} \notag \\
  & + 3 D_{X^{\lambda_1}} \partial_{p_{\lambda_2}} \partial_{p_{\lambda_3}} {\tilde A} \partial_{p_{\lambda_1}} D_{X^{\lambda_2}} D_{X^{\lambda_3}} {\tilde B}
  - \partial_{p_{\lambda_1}} \partial_{p_{\lambda_2}} \partial_{p_{\lambda_3}} {\tilde A} D_{X^{\lambda_1}} D_{X^{\lambda_2}} D_{X^{\lambda_3}} {\tilde B}, \label{eq:wigner8g} \\
  {\cal P}_{{\cal F}^2}({\tilde A}, {\tilde B})
  \equiv & ({\cal F}_{\mu_1 \nu_1} {\cal F}_{\mu_2 \nu_2} \partial_{p_{\mu_1}} \partial_{p_{\mu_2}} {\tilde A} \partial_{p_{\nu_1}} \partial_{p_{\nu_2}} {\tilde B}
  + 4 \partial_{p_{\mu_1}} \partial_{p_{\mu_2}} {\tilde A} {\cal F}_{\mu_1 \nu_1} {\cal F}_{\mu_2 \nu_2} \partial_{p_{\nu_1}} \partial_{p_{\nu_2}} {\tilde B} \notag \\
  & + \partial_{p_{\mu_1}} \partial_{p_{\mu_2}} {\tilde A} \partial_{p_{\nu_1}} \partial_{p_{\nu_2}} {\tilde B} {\cal F}_{\mu_1 \nu_1} {\cal F}_{\mu_2 \nu_2}
  + 4 {\cal F}_{\mu_1 \nu_1} \partial_{p_{\mu_1}} \partial_{p_{\mu_2}} {\tilde A} {\cal F}_{\mu_2 \nu_2} \partial_{p_{\nu_1}} \partial_{p_{\nu_2}} {\tilde B} \notag \\
  & + 4 \partial_{p_{\mu_1}} \partial_{p_{\mu_2}} {\tilde A} {\cal F}_{\mu_1 \nu_1} \partial_{p_{\nu_1}} \partial_{p_{\nu_2}} {\tilde B} {\cal F}_{\mu_2 \nu_2}
  + 2 {\cal F}_{\mu_1 \nu_1} \partial_{p_{\mu_1}} \partial_{p_{\mu_2}} {\tilde A} \partial_{p_{\nu_1}} \partial_{p_{\nu_2}} {\tilde B} {\cal F}_{\mu_2 \nu_2})/4^2, \label{eq:wigner8h} \\
  {\cal P}_{D^2 \ast {\cal F}}({\tilde A}, {\tilde B})
  \equiv & [{\cal F}_{\mu \nu} (D_{X^{\lambda_1}} D_{X^{\lambda_2}} \partial_{p_{\mu}} {\tilde A} \partial_{p_{\lambda_1}} \partial_{p_{\lambda_2}} \partial_{p_{\nu}} {\tilde B}
  - 2 D_{X^{\lambda_1}} \partial_{p_{\lambda_2}} \partial_{p_{\mu}} {\tilde A} \partial_{p_{\lambda_1}} D_{X^{\lambda_2}} \partial_{p_{\nu}} {\tilde B} \notag \\
  & + \partial_{p_{\lambda_1}} \partial_{p_{\lambda_2}} \partial_{p_{\mu}} {\tilde A} D_{X^{\lambda_1}} D_{X^{\lambda_2}} \partial_{p_{\nu}} {\tilde B})
  + 2 (D_{X^{\lambda_1}} D_{X^{\lambda_2}} \partial_{p_{\mu}} {\tilde A} \partial_{p_{\lambda_1}} {\cal F}_{\mu \nu} \partial_{p_{\lambda_2}} \partial_{p_{\nu}} {\tilde B} \notag \\
  & - 2 D_{X^{\lambda_1}} \partial_{p_{\lambda_2}} \partial_{p_{\mu}} {\tilde A} {\cal F}_{\mu \nu} \partial_{p_{\lambda_1}} D_{X^{\lambda_2}} \partial_{p_{\nu}} {\tilde B}
  + \partial_{p_{\lambda_1}} \partial_{p_{\lambda_2}} \partial_{p_{\mu}} {\tilde A} {\cal F}_{\mu \nu} D_{X^{\lambda_1}} D_{X^{\lambda_2}} \partial_{p_{\nu}} {\tilde B}) \notag \\
  & + (D_{X^{\lambda_1}} D_{X^{\lambda_2}} \partial_{p_{\mu}} {\tilde A} \partial_{p_{\lambda_1}} \partial_{p_{\lambda_2}} \partial_{p_{\nu}} {\tilde B}
  - 2 D_{X^{\lambda_1}} \partial_{p_{\lambda_2}} \partial_{p_{\mu}} {\tilde A} \partial_{p_{\lambda_1}} D_{X^{\lambda_2}} \partial_{p_{\nu}} {\tilde B} \notag \\
  & + \partial_{p_{\lambda_1}} \partial_{p_{\lambda_2}} \partial_{p_{\mu}} {\tilde A} D_{X^{\lambda_1}} D_{X^{\lambda_2}} \partial_{p_{\nu}} {\tilde B}) {\cal F}_{\mu \nu}]/4, \label{eq:wigner8i} \\
  {\cal P}_{D \ast D {\cal F}}({\tilde A}, {\tilde B})
  \equiv & \{D_{X^{\lambda_1}} {\cal F}_{\mu \nu} [2 (D_{X^{\lambda_2}} \partial_{p_{\mu}} {\tilde A} \partial_{p_{\lambda_1}} \partial_{p_{\lambda_2}} \partial_{p_{\nu}} {\tilde B}
  - \partial_{p_{\lambda_2}} \partial_{p_{\mu}} {\tilde A} D_{X^{\lambda_2}} \partial_{p_{\lambda_1}} \partial_{p_{\nu}} {\tilde B}) \notag \\
  & + (D_{X^{\lambda_2}} \partial_{p_{\lambda_1}} \partial_{p_{\mu}} {\tilde A} \partial_{p_{\lambda_2}} \partial_{p_{\nu}} {\tilde B}
  - \partial_{p_{\lambda_1}} \partial_{p_{\lambda_2}} \partial_{p_{\mu}} {\tilde A} D_{X^{\lambda_2}} \partial_{p_{\nu}} {\tilde B})] \notag \\
  & + 3 [(D_{X^{\lambda_2}} \partial_{p_{\mu}} {\tilde A} D_{X^{\lambda_1}} {\cal F}_{\mu \nu} \partial_{p_{\lambda_1}} \partial_{p_{\lambda_2}} \partial_{p_{\nu}} {\tilde B}
  - \partial_{p_{\lambda_2}} \partial_{p_{\mu}} {\tilde A} D_{X^{\lambda_1}} {\cal F}_{\mu \nu} D_{X^{\lambda_2}} \partial_{p_{\lambda_1}} \partial_{p_{\nu}} {\tilde B}) \notag \\
  & - (D_{X^{\lambda_2}} \partial_{p_{\lambda_1}} \partial_{p_{\mu}} {\tilde A} D_{X^{\lambda_1}} {\cal F}_{\mu \nu} \partial_{p_{\lambda_2}} \partial_{p_{\nu}} {\tilde B}
  - \partial_{p_{\lambda_1}} \partial_{p_{\lambda_2}} \partial_{p_{\mu}} {\tilde A} D_{X^{\lambda_1}} {\cal F}_{\mu \nu} D_{X^{\lambda_2}} \partial_{p_{\nu}} {\tilde B})] \notag \\
  & - [(D_{X^{\lambda_2}} \partial_{p_{\mu}} {\tilde A} \partial_{p_{\lambda_1}} \partial_{p_{\lambda_2}} \partial_{p_{\nu}} {\tilde B}
  - \partial_{p_{\lambda_2}} \partial_{p_{\mu}} {\tilde A} D_{X^{\lambda_2}} \partial_{p_{\lambda_1}} \partial_{p_{\nu}} {\tilde B}) \notag \\
  & + 2 (D_{X^{\lambda_2}} \partial_{p_{\lambda_1}} \partial_{p_{\mu}} {\tilde A} \partial_{p_{\lambda_2}} \partial_{p_{\nu}} {\tilde B}
  - \partial_{p_{\lambda_1}} \partial_{p_{\lambda_2}} \partial_{p_{\mu}} {\tilde A} D_{X^{\lambda_2}} \partial_{p_{\nu}} {\tilde B})] D_{X^{\lambda_1}} {\cal F}_{\mu \nu}\}/4, \label{eq:wigner8j} \\
  {\cal P}_{D^2 {\cal F}}({\tilde A}, {\tilde B})
  \equiv & \{D_{X^{\lambda_1}} D_{X^{\lambda_2}} {\cal F}_{\mu \nu} [6 \partial_{p_{\mu}} {\tilde A} \partial_{p_{\lambda_1}} \partial_{p_{\lambda_2}} \partial_{p_{\nu}} {\tilde B}
  + 3 (\partial_{p_{\lambda_2}} \partial_{p_{\mu}} {\tilde A} \partial_{p_{\lambda_1}} \partial_{p_{\nu}} {\tilde B}
  + \partial_{p_{\lambda_1}} \partial_{p_{\mu}} {\tilde A} \partial_{p_{\lambda_2}} \partial_{p_{\nu}} {\tilde B}) \notag \\
  & + 2 \partial_{p_{\lambda_1}} \partial_{p_{\lambda_2}} \partial_{p_{\mu}} {\tilde A} \partial_{p_{\nu}} {\tilde B}]
  + [8 \partial_{p_{\mu}} {\tilde A} D_{X^{\lambda_1}} D_{X^{\lambda_2}} {\cal F}_{\mu \nu} \partial_{p_{\lambda_1}} \partial_{p_{\lambda_2}} \partial_{p_{\nu}} {\tilde B}
  - 6 (\partial_{p_{\lambda_2}} \partial_{p_{\mu}} {\tilde A} D_{X^{\lambda_1}} D_{X^{\lambda_2}} {\cal F}_{\mu \nu} \partial_{p_{\lambda_1}} \partial_{p_{\nu}} {\tilde B} \notag \\
  & + \partial_{p_{\lambda_1}} \partial_{p_{\mu}} {\tilde A} D_{X^{\lambda_1}} D_{X^{\lambda_2}} {\cal F}_{\mu \nu} \partial_{p_{\lambda_2}} \partial_{p_{\nu}} {\tilde B})
  + 8 \partial_{p_{\lambda_1}} \partial_{p_{\lambda_2}} \partial_{p_{\mu}} {\tilde A} D_{X^{\lambda_1}} D_{X^{\lambda_2}} {\cal F}_{\mu \nu} \partial_{p_{\nu}} {\tilde B}]
  + [2 \partial_{p_{\mu}} {\tilde A} \partial_{p_{\lambda_1}} \partial_{p_{\lambda_2}} \partial_{p_{\nu}} {\tilde B} \notag \\
  & + 3 (\partial_{p_{\lambda_2}} \partial_{p_{\mu}} {\tilde A} \partial_{p_{\lambda_1}} \partial_{p_{\nu}} {\tilde B}
  + \partial_{p_{\lambda_1}} \partial_{p_{\mu}} {\tilde A} \partial_{p_{\lambda_2}} \partial_{p_{\nu}} {\tilde B})
  + 6 \partial_{p_{\lambda_1}} \partial_{p_{\lambda_2}} \partial_{p_{\mu}} {\tilde A} \partial_{p_{\nu}} {\tilde B}]D_{X^{\lambda_1}} D_{X^{\lambda_2}} {\cal F}_{\mu \nu}\}/16, \label{eq:wigner8k} \\
  {\cal P}_{D^4}({\tilde A}, {\tilde B})
  \equiv & D_{X^{\lambda_1}} D_{X^{\lambda_2}} D_{X^{\lambda_3}} D_{X^{\lambda_4}} {\tilde A} \partial_{p_{\lambda_1}} \partial_{p_{\lambda_2}} \partial_{p_{\lambda_3}} \partial_{p_{\lambda_4}}{\tilde B}
  - 4 D_{X^{\lambda_1}} D_{X^{\lambda_2}} D_{X^{\lambda_3}} \partial_{p_{\lambda_4}} {\tilde A} \partial_{p_{\lambda_1}} \partial_{p_{\lambda_2}} \partial_{p_{\lambda_4}} D_{X^{\lambda_4}} {\tilde B} \notag \\
  & + 6 D_{X^{\lambda_1}} D_{X^{\lambda_2}} \partial_{p_{\lambda_3}} \partial_{p_{\lambda_4}} {\tilde A} \partial_{p_{\lambda_1}} \partial_{p_{\lambda_2}} D_{X^{\lambda_3}} D_{X^{\lambda_4}} {\tilde B}
  - 4 D_{X^{\lambda_1}} \partial_{p_{\lambda_2}} \partial_{p_{\lambda_3}} \partial_{p_{\lambda_4}} {\tilde A} \partial_{p_{\lambda_1}} D_{X^{\lambda_2}} D_{X^{\lambda_3}} D_{X^{\lambda_4}} {\tilde B} \notag \\
  & + \partial_{p_{\lambda_1}} \partial_{p_{\lambda_2}} \partial_{p_{\lambda_3}} \partial_{p_{\lambda_4}} {\tilde A} D_{X^{\lambda_1}} D_{X^{\lambda_2}} D_{X^{\lambda_3}} D_{X^{\lambda_4}} {\tilde B}. \label{eq:wigner8l}
\end{align} \label{eq:wigner8}\end{subequations}
The arguments $X, p$ are omitted.
Equations~\eqref{eq:wigner8f}, \eqref{eq:wigner8g}, and \eqref{eq:wigner8i}-\eqref{eq:wigner8l} are not written in Eq.~\eqref{eq:moyal2}.

\section{Equivalence of the left and right Dyson equations} \label{app:equivalence}
Equations~\eqref{eq:grad2}, \eqref{eq:first2}, and \eqref{eq:second2} are derived from the left Dyson equation.
From the right Dyson equation,
\begin{equation}
  {\tilde G} \star ({\tc L} - {\tilde \Sigma})
  = 1, \label{eq:dyson2}
\end{equation}
we also obtain
\begin{subequations} \begin{align}
  {\tilde G}_P {\tilde G}_0^{-1}
  = & {\tilde G}_0 {\tilde \Sigma}_P - i {\cal P}_P({\tilde G}_0, {\tilde G}_0^{-1}), \label{eq:grad3a} \\
  {\tilde G}_{P \ast Q} {\tilde G}_0^{-1}
  = & {\tilde G}_0 {\tilde \Sigma}_{P \ast Q} - i^2 {\cal P}_{P \ast Q}({\tilde G}_0, {\tilde G}_0^{-1})
  + [{\tilde G}_P {\tilde \Sigma}_Q + i {\cal P}_P({\tilde G}_0, {\tilde \Sigma}_Q) - i {\cal P}_P({\tilde G}_Q, {\tilde G}_0^{-1}) + (P \leftrightarrow Q)], \label{eq:grad3b}
\end{align} \label{eq:grad3}\end{subequations}
\begin{subequations} \begin{align}
  G_P^{\rm R} (G_0^{\rm R})^{-1}
  = & G_0^{\rm R} \Sigma_P^{\rm R} - i \eta_P^{IJ} \partial_I G_0^{\rm R} \partial_J (G_0^{\rm R})^{-1}, \label{eq:first3a} \\
  G_P^{\rm A} (G_0^{\rm A})^{-1}
  = & G_0^{\rm A} \Sigma_P^{\rm A} - i \eta_P^{IJ} \partial_I G_0^{\rm A} \partial_J (G_0^{\rm A})^{-1}, \label{eq:first3b} \\
  G_P^{< (1)} (G_0^{\rm A})^{-1}
  = & G_0^{\rm R} \Sigma_P^{< (1)}
  + i \eta_P^{I p_0} \{\partial_I G_0^{\rm R} [(G_0^{\rm R})^{-1} - (G_0^{\rm A})^{-1}] - (G_0^{\rm R} - G_0^{\rm A}) \partial_I (G_0^{\rm A})^{-1}\}, \label{eq:first3c}
\end{align} \label{eq:first3}\end{subequations}
and
\begin{subequations} \begin{align}
  G_{P \ast Q}^{\rm R} (G_0^{\rm R})^{-1}
  = & G_0^{\rm R} \Sigma_{P \ast Q}^{\rm R}
  + [G_P^{\rm R} \Sigma_Q^{\rm R} + i \eta_P^{IJ} \partial_I G_0^{\rm R} \partial_J \Sigma_Q^{\rm R} - i \eta_P^{IJ} \partial_I G_Q^{\rm R} \partial_J (G_0^{\rm R})^{-1} + (P \leftrightarrow Q)] \notag \\
  & + \eta_P^{IJ} \eta_Q^{KL} \partial_I \partial_K G_0^{\rm R} \partial_J \partial_L (G_0^{\rm R})^{-1}, \label{eq:second3a} \\
  G_{P \ast Q}^{\rm A} (G_0^{\rm A})^{-1}
  = & G_0^{\rm A} \Sigma_{P \ast Q}^{\rm A}
  + [G_P^{\rm A} \Sigma_Q^{\rm A} + i \eta_P^{IJ} \partial_I G_0^{\rm A} \partial_J \Sigma_Q^{\rm A} - i \eta_P^{IJ} \partial_I G_Q^{\rm A} \partial_J (G_0^{\rm A})^{-1} + (P \leftrightarrow Q)] \notag \\
  & + \eta_P^{IJ} \eta_Q^{KL} \partial_I \partial_K G_0^{\rm A} \partial_J \partial_L (G_0^{\rm A})^{-1}, \label{eq:second3b} \\
  G_{P \ast Q}^{< (1)} (G_0^{\rm A})^{-1}
  = & G_0^{\rm R} \Sigma_{P \ast Q}^{< (1)} + \left(G_P^{< (1)} \Sigma_Q^{\rm A} + G_P^{\rm R} \Sigma_Q^{< (1)}
  + i \eta_P^{IJ} \partial_I G_0^{\rm R} \partial_J \Sigma_Q^{< (1)}
  - i \eta_P^{I p_0} [\partial_I G_0^{\rm R} (\Sigma_Q^{\rm R} - \Sigma_Q^{\rm A}) - (G_0^{\rm R} - G_0^{\rm A}) \partial_I \Sigma_Q^{\rm A}] \right.\notag \\
  & - i \eta_P^{IJ} \partial_I G_Q^{< (1)} \partial_J (G_0^{\rm A})^{-1}
  + i \eta_P^{I p_0} \{\partial_I G_Q^{\rm R} [(G_0^{\rm R})^{-1} - (G_0^{\rm A})^{-1}] - (G_Q^{\rm R} - G_Q^{\rm A}) \partial_I (G_0^{\rm A})^{-1}\} \notag \\
  & \left.- \eta_P^{I p_0} \eta_Q^{KL} \{\partial_I \partial_K G_0^{\rm R} \partial_L [(G_0^{\rm R})^{-1} - (G_0^{\rm A})^{-1}] + \partial_L (G_0^{\rm R} - G_0^{\rm A}) \partial_I \partial_K (G_0^{\rm A})^{-1}\}
  + (P \leftrightarrow Q)\right), \label{eq:second3c} \\
  G_{P \ast Q}^{< (2)} (G_0^{\rm A})^{-1}
  = & G_0^{\rm R} \Sigma_{P \ast Q}^{< (2)}
  + [- i \eta_P^{I p_0} \partial_I G_0^{\rm R} \Sigma_Q^{< (1)} - i \eta_P^{I p_0} G_Q^{< (1)} \partial_I (G_0^{\rm A})^{-1} + (P \leftrightarrow Q)] \notag \\
  & + \eta_P^{I p_0} \eta_Q^{K p_0} \{\partial_I \partial_K G_0^{\rm R} [(G_0^{\rm R})^{-1} - (G_0^{\rm A})^{-1}] + (G_0^{\rm R} - G_0^{\rm A}) \partial_I \partial_K (G_0^{\rm A})^{-1}\}. \label{eq:second3d}
\end{align} \label{eq:second3}\end{subequations}
Thus, the left and right Dyson equations seem different from each other but in fact are equivalent.
Both equations lead to
\begin{subequations} \begin{align}
  G_P^{\rm R}
  = & G_0^{\rm R} \Sigma_P^{\rm R} G_0^{\rm R} + i \eta_P^{IJ} G_0^{\rm R} \partial_I (G_0^{\rm R})^{-1} G_0^{\rm R} \partial_J (G_0^{\rm R})^{-1} G_0^{\rm R}, \label{eq:first4a} \\
  G_P^{\rm A}
  = & G_0^{\rm A} \Sigma_P^{\rm A} G_0^{\rm A} + i \eta_P^{IJ} G_0^{\rm A} \partial_I (G_0^{\rm A})^{-1} G_0^{\rm A} \partial_J (G_0^{\rm A})^{-1} G_0^{\rm A}, \label{eq:first4b} \\
  G_P^{< (1)}
  = & G_0^{\rm R} \Sigma_P^{< (1)} G_0^{\rm A} - i \eta_P^{I p_0} \{G_0^{\rm R} \partial_I [(G_0^{\rm R})^{-1} + (G_0^{\rm A})^{-1}] G_0^{\rm A} + \partial_I (G_0^{\rm R} + G_0^{\rm A})\}. \label{eq:first4c}
\end{align} \label{eq:first4}\end{subequations}
and
\begin{subequations} \begin{align}
  G_{P \ast Q}^{\rm R}
  = & G_0^{\rm R} \Sigma_{P \ast Q}^{\rm R} G_0^{\rm R}
  + \{G_0^{\rm R} \Sigma_Q^{\rm R} G_0^{\rm R} \Sigma_P^{\rm R} G_0^{\rm R}
  + i \eta_P^{IJ} G_0^{\rm R} [-\partial_I \Sigma_Q^{\rm R} G_0^{\rm R} \partial_J (G_0^{\rm R})^{-1} - \partial_I (G_0^{\rm R})^{-1} G_0^{\rm R} \partial_J \Sigma_Q^{\rm R} \notag \\
  & + \Sigma_Q^{\rm R} G_0^{\rm R} \partial_I (G_0^{\rm R})^{-1} G_0^{\rm R} \partial_J (G_0^{\rm R})^{-1} + \partial_I (G_0^{\rm R})^{-1} G_0^{\rm R} \Sigma_Q^{\rm R} G_0^{\rm R} \partial_J G_0^{\rm R}
  + G_0^{\rm R} \partial_I (G_0^{\rm R})^{-1} G_0^{\rm R} \partial_J (G_0^{\rm R})^{-1} G_0^{\rm R} \Sigma_Q^{\rm R}] G_0^{\rm R} \notag \\
  & + (P \leftrightarrow Q)\}
  + \eta_P^{IJ} \eta_Q^{KL} G_0^{\rm R} \{-\partial_I \partial_K (G_0^{\rm R})^{-1} G_0^{\rm R} \partial_J \partial_L (G_0^{\rm R})^{-1} \notag \\
  & - \partial_J (G_0^{\rm R})^{-1} G_0^{\rm R} \partial_I \partial_K (G_0^{\rm R})^{-1} G_0^{\rm R} \partial_L (G_0^{\rm R})^{-1}
  - \partial_L (G_0^{\rm R})^{-1} G_0^{\rm R} \partial_I \partial_K (G_0^{\rm R})^{-1} G_0^{\rm R} \partial_J (G_0^{\rm R})^{-1} \notag \\
  & + \partial_I \partial_K (G_0^{\rm R})^{-1} G_0^{\rm R} [\partial_J (G_0^{\rm R})^{-1} G_0^{\rm R} \partial_L (G_0^{\rm R})^{-1} + \partial_L (G_0^{\rm R})^{-1} G_0^{\rm R} \partial_J (G_0^{\rm R})^{-1}] \notag \\
  & + [\partial_J (G_0^{\rm R})^{-1} G_0^{\rm R} \partial_L (G_0^{\rm R})^{-1} + \partial_L (G_0^{\rm R})^{-1} G_0^{\rm R} \partial_J (G_0^{\rm R})^{-1}] G_0^{\rm R} \partial_I \partial_K (G_0^{\rm R})^{-1} \notag \\
  & - \partial_I (G_0^{\rm R})^{-1} G_0^{\rm R} \partial_J (G_0^{\rm R})^{-1} G_0^{\rm R} \partial_K (G_0^{\rm R})^{-1} G_0^{\rm R} \partial_L (G_0^{\rm R})^{-1}
  - \partial_I (G_0^{\rm R})^{-1} G_0^{\rm R} \partial_K (G_0^{\rm R})^{-1} G_0^{\rm R} \partial_J (G_0^{\rm R})^{-1} G_0^{\rm R} \partial_L (G_0^{\rm R})^{-1} \notag \\
  & - \partial_I (G_0^{\rm R})^{-1} G_0^{\rm R} \partial_K (G_0^{\rm R})^{-1} G_0^{\rm R} \partial_L (G_0^{\rm R})^{-1} G_0^{\rm R} \partial_J (G_0^{\rm R})^{-1}
  - \partial_K (G_0^{\rm R})^{-1} G_0^{\rm R} \partial_L (G_0^{\rm R})^{-1} G_0^{\rm R} \partial_I (G_0^{\rm R})^{-1} G_0^{\rm R} \partial_J (G_0^{\rm R})^{-1} \notag \\
  & - \partial_K (G_0^{\rm R})^{-1} G_0^{\rm R} \partial_I (G_0^{\rm R})^{-1} G_0^{\rm R} \partial_L (G_0^{\rm R})^{-1} G_0^{\rm R} \partial_J (G_0^{\rm R})^{-1}
  - \partial_K (G_0^{\rm R})^{-1} G_0^{\rm R} \partial_I (G_0^{\rm R})^{-1} G_0^{\rm R} \partial_J (G_0^{\rm R})^{-1} G_0^{\rm R} \partial_L (G_0^{\rm R})^{-1}\} \notag \\
  & \times G_0^{\rm R}, \label{eq:second4a} \\
  G_{P \ast Q}^{\rm A}
  = & G_0^{\rm A} \Sigma_{P \ast Q}^{\rm A} G_0^{\rm A}
  + \{G_0^{\rm A} \Sigma_Q^{\rm A} G_0^{\rm A} \Sigma_P^{\rm A} G_0^{\rm A}
  + i \eta_P^{IJ} G_0^{\rm A} [-\partial_I \Sigma_Q^{\rm A} G_0^{\rm A} \partial_J (G_0^{\rm A})^{-1} - \partial_I (G_0^{\rm A})^{-1} G_0^{\rm A} \partial_J \Sigma_Q^{\rm A} \notag \\
  & + \Sigma_Q^{\rm A} G_0^{\rm A} \partial_I (G_0^{\rm A})^{-1} G_0^{\rm A} \partial_J (G_0^{\rm A})^{-1} + \partial_I (G_0^{\rm A})^{-1} G_0^{\rm A} \Sigma_Q^{\rm A} G_0^{\rm A} \partial_J G_0^{\rm A}
  + G_0^{\rm A} \partial_I (G_0^{\rm A})^{-1} G_0^{\rm A} \partial_J (G_0^{\rm A})^{-1} G_0^{\rm A} \Sigma_Q^{\rm A}] G_0^{\rm A} \notag \\
  & + (P \leftrightarrow Q)\}
  + \eta_P^{IJ} \eta_Q^{KL} G_0^{\rm A} \{-\partial_I \partial_K (G_0^{\rm A})^{-1} G_0^{\rm A} \partial_J \partial_L (G_0^{\rm A})^{-1} \notag \\
  & - \partial_J (G_0^{\rm A})^{-1} G_0^{\rm A} \partial_I \partial_K (G_0^{\rm A})^{-1} G_0^{\rm A} \partial_L (G_0^{\rm A})^{-1}
  - \partial_L (G_0^{\rm A})^{-1} G_0^{\rm A} \partial_I \partial_K (G_0^{\rm A})^{-1} G_0^{\rm A} \partial_J (G_0^{\rm A})^{-1} \notag \\
  & + \partial_I \partial_K (G_0^{\rm A})^{-1} G_0^{\rm A} [\partial_J (G_0^{\rm A})^{-1} G_0^{\rm A} \partial_L (G_0^{\rm A})^{-1} + \partial_L (G_0^{\rm A})^{-1} G_0^{\rm A} \partial_J (G_0^{\rm A})^{-1}] \notag \\
  & + [\partial_J (G_0^{\rm A})^{-1} G_0^{\rm A} \partial_L (G_0^{\rm A})^{-1} + \partial_L (G_0^{\rm A})^{-1} G_0^{\rm A} \partial_J (G_0^{\rm A})^{-1}] G_0^{\rm A} \partial_I \partial_K (G_0^{\rm A})^{-1} \notag \\
  & - \partial_I (G_0^{\rm A})^{-1} G_0^{\rm A} \partial_J (G_0^{\rm A})^{-1} G_0^{\rm A} \partial_K (G_0^{\rm A})^{-1} G_0^{\rm A} \partial_L (G_0^{\rm A})^{-1}
  - \partial_I (G_0^{\rm A})^{-1} G_0^{\rm A} \partial_K (G_0^{\rm A})^{-1} G_0^{\rm A} \partial_J (G_0^{\rm A})^{-1} G_0^{\rm A} \partial_L (G_0^{\rm A})^{-1} \notag \\
  & - \partial_I (G_0^{\rm A})^{-1} G_0^{\rm A} \partial_K (G_0^{\rm A})^{-1} G_0^{\rm A} \partial_L (G_0^{\rm A})^{-1} G_0^{\rm A} \partial_J (G_0^{\rm A})^{-1}
  - \partial_K (G_0^{\rm A})^{-1} G_0^{\rm A} \partial_L (G_0^{\rm A})^{-1} G_0^{\rm A} \partial_I (G_0^{\rm A})^{-1} G_0^{\rm A} \partial_J (G_0^{\rm A})^{-1} \notag \\
  & - \partial_K (G_0^{\rm A})^{-1} G_0^{\rm A} \partial_I (G_0^{\rm A})^{-1} G_0^{\rm A} \partial_L (G_0^{\rm A})^{-1} G_0^{\rm A} \partial_J (G_0^{\rm A})^{-1}
  - \partial_K (G_0^{\rm A})^{-1} G_0^{\rm A} \partial_I (G_0^{\rm A})^{-1} G_0^{\rm A} \partial_J (G_0^{\rm A})^{-1} G_0^{\rm A} \partial_L (G_0^{\rm A})^{-1}\} \notag \\
  & \times G_0^{\rm A}, \label{eq:second4b} \\
  G_{P \ast Q}^{< (1)}
  = & G_0^{\rm R} \Sigma_{P \ast Q}^{< (1)} G_0^{\rm A}
  + \left(G_0^{\rm R} (\Sigma_Q^{\rm R} G_0^{\rm R} \Sigma_P^{< (1)} + \Sigma_Q^{< (1)} G_0^{\rm A} \Sigma_P^{\rm A}) G_0^{\rm A}
  + i \eta_P^{IJ} G_0^{\rm R} [-\partial_I \Sigma_Q^{< (1)} G_0^{\rm A} \partial_J (G_0^{\rm A})^{-1} - \partial_I (G_0^{\rm R})^{-1} G_0^{\rm R} \partial_J \Sigma_Q^{< (1)}\right. \notag \\
  & + \Sigma_Q^{< (1)} G_0^{\rm A} \partial_I (G_0^{\rm A})^{-1} G_0^{\rm A} \partial_J (G_0^{\rm A})^{-1} + \partial_I (G_0^{\rm R})^{-1} G_0^{\rm R} \partial_J (G_0^{\rm R})^{-1} G_0^{\rm R} \Sigma_Q^{< (1)}
  + \partial_I (G_0^{\rm R})^{-1} G_0^{\rm R} \Sigma_Q^{< (1)} G_0^{\rm A} \partial_J (G_0^{\rm A})^{-1}] G_0^{\rm A} \notag \\
  & + i \eta_P^{I p_0} G_0^{\rm R} \{\partial_I (\Sigma_Q^{\rm R} + \Sigma_Q^{\rm A})
  - \Sigma_Q^{\rm R} G_0^{\rm R} \partial_I [(G_0^{\rm R})^{-1} + (G_0^{\rm A})^{-1}] - \partial_I [(G_0^{\rm R})^{-1} + (G_0^{\rm A})^{-1}] G_0^{\rm A} \Sigma_Q^{\rm A}\} G_0^{\rm A} \notag \\
  & + \eta_P^{I p_0} \eta_Q^{KL} G_0^{\rm R}
  \{-\partial_I \partial_K [(G_0^{\rm R})^{-1} + (G_0^{\rm A})^{-1}] G_0^{\rm A} \partial_L (G_0^{\rm A})^{-1} - \partial_K (G_0^{\rm R})^{-1} G_0^{\rm R} \partial_I \partial_L [(G_0^{\rm R})^{-1} + (G_0^{\rm A})^{-1}] \notag \\
  & + \partial_K (G_0^{\rm R})^{-1} G_0^{\rm R} \partial_I [(G_0^{\rm R})^{-1} + (G_0^{\rm A})^{-1}] G_0^{\rm A} \partial_L (G_0^{\rm A})^{-1}
  + \partial_I [(G_0^{\rm R})^{-1} + (G_0^{\rm A})^{-1}] G_0^{\rm A} \partial_K (G_0^{\rm A})^{-1} G_0^{\rm A} \partial_L (G_0^{\rm A})^{-1} \notag \\
  & \left.+ \partial_K (G_0^{\rm R})^{-1} G_0^{\rm R} \partial_L (G_0^{\rm R})^{-1} G_0^{\rm R} \partial_I [(G_0^{\rm R})^{-1} + (G_0^{\rm A})^{-1}]\} G_0^{\rm A}
  - i \eta_P^{I p_0} \partial_I (G_Q^{\rm R} + G_Q^{\rm A})+ (P \leftrightarrow Q)\right), \label{eq:second4c} \\
  G_{P \ast Q}^{< (2)}
  = & G_0^{\rm R} \Sigma_{P \ast Q}^{< (2)} G_0^{\rm A}
  + \{i \eta_P^{I p_0} G_0^{\rm R} [\partial_I (G_0^{\rm R})^{-1} G_0^{\rm R} \Sigma_Q^{< (1)} - \Sigma_Q^{< (1)} G_0^{\rm A} \partial_I (G_0^{\rm A})^{-1}] G_0^{\rm A} + (P \leftrightarrow Q)\} \notag \\
  & + \eta_P^{I p_0} \eta_Q^{K p_0} \left(G_0^{\rm R} \{-\partial_I \partial_K [(G_0^{\rm R})^{-1} - (G_0^{\rm A})^{-1}]
  - \partial_I [(G_0^{\rm R})^{-1} + (G_0^{\rm A})^{-1}] G_0^{\rm A} \partial_K (G_0^{\rm A})^{-1}\right. \notag \\
  & - \partial_K [(G_0^{\rm R})^{-1} + (G_0^{\rm A})^{-1}] G_0^{\rm A} \partial_I (G_0^{\rm A})^{-1}
  + \partial_I (G_0^{\rm R})^{-1} G_0^{\rm R} \partial_K [(G_0^{\rm R})^{-1} + (G_0^{\rm A})^{-1}] \notag \\
  & \left.+ \partial_K (G_0^{\rm R})^{-1} G_0^{\rm R} \partial_I [(G_0^{\rm R})^{-1} + (G_0^{\rm A})^{-1}]\} G_0^{\rm A} + \partial_I \partial_K (G_0^{\rm R} - G_0^{\rm A})\right). \label{eq:second4d}
\end{align} \label{eq:second4}\end{subequations}

\section{Momentum integrals in Eqs.~\eqref{eq:ferroint1} and \eqref{eq:ferroint2}} \label{app:integral}
In this Appendix, we give the explicit forms of the momentum integrals defined above.
Here we introduce
\begin{subequations} \begin{align}
  D_0(E)
  \equiv & -\int_0^{\Lambda} \frac{\sqrt{x} {\rm d} x}{\pi} \frac{1}{E - x}
  = 2 [\Lambda^{1/2} - E^{1/2} \tanh^{-1} (\Lambda/E)^{1/2}]/\pi, \label{eq:ferroint3a} \\
  D_1(E)
  = & \int_0^{\Lambda} \frac{\sqrt{x} {\rm d} x}{\pi} \frac{1}{(E - x)^2}
  = [\Lambda^{1/2}/(\Lambda - E) - E^{-1/2} \tanh^{-1} (\Lambda/E)^{1/2}]/\pi, \label{eq:ferroint3b}
\end{align} \label{eq:ferroint3}\end{subequations}
and $E_{\pm}^{\rm R}(\xi) \equiv \xi^{\rm R}(\xi) \pm J^{\rm R}(\xi)$.
Equation~\eqref{eq:ferroint1} is given by
\begin{subequations} \begin{align}
  I_{01}^{\rm R}(\xi)
  = & [D_0(E_+^{\rm R}(\xi)) - D_0(E_-^{\rm R}(\xi))]/2, \label{eq:ferroint4a} \\
  I_{11}^{\rm R}(\xi)
  = & -[D_0(E_+^{\rm R}(\xi)) + D_0(E_-^{\rm R}(\xi))]/2, \label{eq:ferroint4b} \\
  I_{02}^{\rm R}(\xi)
  = & -[D_0(E_+^{\rm R}(\xi)) - D_0(E_-^{\rm R}(\xi)) - J^{\rm R}(\xi) D_1(E_+^{\rm R}(\xi)) - J^{\rm R}(\xi) D_1(E_-^{\rm R}(\xi))]/4, \label{eq:ferroint4c} \\
  I_{12}^{\rm R}(\xi)
  = & -[J^{\rm R}(\xi) D_1(E_+^{\rm R}(\xi)) - J^{\rm R}(\xi) D_1(E_-^{\rm R}(\xi))]/4, \label{eq:ferroint4d} \\
  I_{22}^{\rm R}(\xi)
  = & [D_0(E_+^{\rm R}(\xi)) - D_0(E_-^{\rm R}(\xi)) + J^{\rm R}(\xi) D_1(E_+^{\rm R}(\xi)) + J^{\rm R}(\xi) D_1(E_-^{\rm R}(\xi))]/4, \label{eq:ferroint4e}
\end{align} \label{eq:ferroint4}\end{subequations}
and Eq.~\eqref{eq:ferroint2} by
\begin{subequations} \begin{align}
  J_1^{< (1)}(\xi)
  = & \frac{1}{2} \left[\frac{D_0(E_+^{\rm R}(\xi))}{E_+^{\rm R}(\xi) - E_-^{\rm A}(\xi)} + \frac{D_0(E_-^{\rm R}(\xi))}{E_-^{\rm R}(\xi) - E_+^{\rm A}(\xi)}\right] + \cc, \label{eq:ferroint5a} \\
  J_2^{< (1)}(\xi)
  = & \frac{i}{2} \left[\frac{D_0(E_+^{\rm R}(\xi))}{E_+^{\rm R}(\xi) - E_-^{\rm A}(\xi)} - \frac{D_0(E_-^{\rm R}(\xi))}{E_-^{\rm R}(\xi) - E_+^{\rm A}(\xi)}\right] + \cc, \label{eq:ferroint5b} \\
  J_3^{< (1)}(\xi)
  = & J_2^{< (1)}(\xi) - \frac{1}{3} [i J^{\rm R}(\xi) - i J^{\rm A}(\xi)] \notag \\
  & \times \left\{\frac{E_+^{\rm R}(\xi) D_0(E_+^{\rm R}(\xi))}{J^{\rm R}(\xi) [E_+^{\rm R}(\xi) - E_+^{\rm A}(\xi)] [E_+^{\rm R}(\xi) - E_-^{\rm A}(\xi)]}
  - \frac{E_-^{\rm R}(\xi) D_0(E_-^{\rm R}(\xi))}{J^{\rm R}(\xi) [E_-^{\rm R}(\xi) - E_-^{\rm A}(\xi)] [E_-^{\rm R}(\xi) - E_+^{\rm A}(\xi)]} + \cc\right\}, \label{eq:ferroint5c} \\
  J_4^{< (1)}(\xi)
  = & \frac{1}{3}
  \left\{\frac{[E_+^{\rm R}(\xi)]^2 - |\xi^{\rm R}(\xi)|^2 + 2 |J^{\rm R}(\xi)|^2}{J^{\rm R}(\xi) [E_+^{\rm R}(\xi) - E_+^{\rm A}(\xi)] [E_+^{\rm R}(\xi) - E_-^{\rm A}(\xi)]} D_0(E_+^{\rm R}(\xi))\right. \notag \\
  & - \frac{[E_-^{\rm R}(\xi)]^2 - |\xi^{\rm R}(\xi)|^2 + 2 |J^{\rm R}(\xi)|^2}{J^{\rm R}(\xi) [E_-^{\rm R}(\xi) - E_-^{\rm A}(\xi)] [E_-^{\rm R}(\xi) - E_+^{\rm A}(\xi)]} D_0(E_-^{\rm R}(\xi)) \notag \\
  & \left.- \frac{E_+^{\rm R}(\xi)}{E_+^{\rm R}(\xi) - E_+^{\rm A}(\xi)} D_1(E_+^{\rm R}(\xi))
  - \frac{E_-^{\rm R}(\xi)}{E_-^{\rm R}(\xi) - E_-^{\rm A}(\xi)} D_1(E_-^{\rm R}(\xi))\right\} + \cc \label{eq:ferroint5d}
\end{align} \label{eq:ferroint5}\end{subequations}
\end{widetext}
%
\end{document}